\documentclass{article}
\usepackage[utf8]{inputenc}
\usepackage[english]{babel} 
\usepackage{amsmath,amssymb,amsthm} 
\usepackage[margin=2cm]{geometry}
\usepackage{upgreek}
\usepackage{esint}
\usepackage[colorlinks]{hyperref}
\usepackage{todonotes}
\usepackage{enumerate}
\usepackage{framed}
\usepackage{pgfplots}
\usepackage{float}
\usepackage{tikz,tikz-cd}
\usepackage[all]{xy}
\usepackage{authblk}
\usepackage{caption}
\usepackage{subcaption}

\begin{document}

\title{Data-driven stochastic Lie transport modelling of the 2D Euler equations}
\date{}                                           
\author[1]{Sagy Ephrati}
\author[1,2]{Paolo Cifani}
\author[1]{Erwin Luesink}
\author[1,3]{Bernard J. Geurts}
\affil[1]{Multiscale Modeling and Simulation, Faculty EEMCS, University of Twente, P.O. Box 217, 7500 AE Enschede, The Netherlands}
\affil[2]{Gran Sasso Science Institute, Viale F. Crispi, 7 67100 L’Aquila, Italy}
\affil[3]{Multiscale Energy Physics, CCER, Faculty Applied Physics, Eindhoven University of Technology, P.O. Box 213, 5600 MB Eindhoven, The Netherlands}
\maketitle

\section*{Abstract}
In this paper, we propose and assess several stochastic parametrizations for data-driven modelling of the two-dimensional Euler equations using coarse-grid SPDEs. The framework of Stochastic Advection by Lie Transport (SALT) \cite{cotter2019numerically} is employed to define a stochastic forcing that is decomposed in terms of a deterministic basis (empirical orthogonal functions, EOFs) multiplied by temporal traces, here regarded as stochastic processes. The EOFs are obtained from a fine-grid data set and are defined in conjunction with corresponding deterministic time series. We construct stochastic processes that mimic properties of the measured time series. In particular, the processes are defined such that the underlying probability density functions (pdfs) or the estimated correlation time of the time series are retained. These stochastic models are compared to stochastic forcing based on Gaussian noise, which does not use any information of the time series. We perform uncertainty quantification tests and compare stochastic ensembles in terms of mean and spread. Reduced uncertainty is observed for the developed models. On short timescales, such as those used for data assimilation \cite{cotter2020data}, the stochastic models show a reduced ensemble mean error and a reduced spread. Particularly, using estimated pdfs yields stochastic ensembles which rarely fail to capture the reference solution on small time scales, whereas introducing correlation into the stochastic models improves the quality of the coarse-grid predictions with respect to Gaussian noise.

\section{\label{sec:Introduction}Introduction}
A major challenge in geophysical and observational sciences is the representation and quantification of uncertainty in numerical predictions. Uncertainty stems from various sources, most relevantly from incomplete inclusion of all relevant physical mechanisms in the models and uncertainty in the initial and boundary conditions \cite{palmer2000predicting}. Important models for geophysical fluid dynamics, such as the two-dimensional Euler equations, quasi-geostrophic equations or rotating shallow water equations are derived from the three-dimensional Navier-Stokes equations. A sequence of simplifying assumptions is applied in order to reduce the complexity of the model to a more manageable level, while retaining main flow physics \cite{zeitlin2018geophysical}. Stochastic extensions to these models have also been derived \cite{holm2021stochastic}. These approximate models are nevertheless are rich in dynamics and contain a wide range of spatial and temporal scales. Numerically resolving the entire spectrum of scales is often not computationally feasible, meaning that either the complexity of the model should be reduced even further such that the resulting model is simple enough the be solvable numerically, or the complex model is represented on a coarse computational grid and unresolved scales are replaced by a sub-grid model. The latter option may be combined with stochastic forcing, which provides an effective way to represent unresolved scales in numerical simulations \cite{buizza1999stochastic}. The use of stochasticity as a means to represent the unresolved scales was introduced by \cite{majda2001mathematical} and serves to restore some of the missing small-scale dynamics and at the same time probes an ensemble of solutions and hence also investigates uncertainty. In this paper, we embrace these ideas and develop and assess stochastic data-driven models for the two-dimensional Euler equations on the unit square.

    
Data-driven stochastic models in dynamical systems have been studied actively in recent years. For weather and climate models, stochasticity was used as a tool to represent uncertainty in initial conditions and in the model, as shown in \cite{palmer2019stochastic}. A commonly used example to illustrate the data-driven stochastic approach is the Lorenz '96 (L96) system, introduced in \cite{lorenz1996predictability} and originally proposed as a simplified model of the atmosphere that incorporates interactions between slow and fast scales. Data of the unresolved sub-grid scales may serve to construct a data-informed stochastic model. Examples are given in \cite{arnold2013stochastic} where sub-grid features are modeled using different types of noise including additive, multiplicative and state-dependent noise. This study established that stochastic parametrizations could accurately account for modelling error, with a considerably improved forecasting skill when temporal correlation was included in the noise. The correlated noise was modeled as a one-step autoregressive model with parameters fitted from data. Alternative ideas such as stochastic parametrization based on Markov chains inferred from data are presented by \cite{crommelin2008subgrid}, where unresolved processes are represented as stochastic processes dependent on the state of the resolved variables and an assumed probability density. Using this approach, good agreement was found for the probability density functions and autocorrelation functions of resolved state variables. 

Data-driven machine learning has also been adopted to represent small-scale dynamics for a large range of parameters \cite{gagne2020machine}. It was found that several configurations of machine learning accurately reconstructed spatio-temporal correlations of the original system. These methods are not limited to simplified models such as the L96 system, but have also been successfully applied to more complete geophysical models. Examples include oceanic flows as considered in \cite{bolton2019applications} and atmospheric processes as investigated in \cite{o2018using}. Both studies obtain a parametrization using machine learning based on off-line computed high-resolution model output. This machine learning approach could accurately predict the relation between resolved and unresolved turbulent processes, although a reliable generalization is principally not guaranteed. Here, we follow another data-driven `offline/online' route and express the differences between a fully resolved model and a coarsened model in terms of a converging series of empirical orthogonal functions (EOFs) and introduce explicit forcing to update the coarsened model to high accuracy. This direct forcing strategy can also be extended to structure-preserving stochastic models as will be clarified below. 

In the seminal work \cite{holm2015variational}, stochastic partial differential equations are derived for fluid dynamics by means of a variational principle. As a result, the solution of the SPDE is compliant with the geometry of the underlying equations. This means that conservation laws are maintained under the inclusion of stochastic perturbations. This approach goes by the name of stochastic advection by Lie transport (SALT). In SALT, spatial correlations of observational data can be used to model the unresolved scales in a numerical simulation. The spatial correlations can be decomposed into EOFs \cite{lumley1967structure, hannachi2007empirical}. These are coupled to noise generated from stochastic processes in a separate modeling step. Together, these terms constitute a stochastic forcing term for the coarse PDE, which models unresolved scales.
    
The SALT approach finds meaningful applications within geophysical fluid dynamics, since these models are directly based on a variational point of view. To illustrate the SALT approach, \cite{cotter2019numerically} apply it to the two-dimensional Euler equations. In this study a fine-grid simulation is performed from which the Lagrangian trajectories and the corresponding trajectories given by filtering the velocity field. The difference between these trajectories is a measure of the unresolved scales to which the EOF decomposition is applied to form an optimal basis for this term. Subsequently, a coarse SPDE is constructed according to SALT where the amplitude of the EOF basis is modeled as a decorrelated stochastic Gaussian process. It is shown that an ensemble of stochastically forced flows captures the mean values of the true solution over considerable time intervals. In a follow-up study \cite{cotter2020particle}, a particle filter was added to the SALT two-dimensional Euler equations and data assimilation was motivated this way. It was demonstrated that significant model reduction is possible, reducing the number of degrees of freedom by two orders of magnitude without losing reliability of the results. Similar studies on the quasi-geostrophic equations have been done \cite{cotter2018modelling}, with a focus on data assimilation \cite{cotter2020data}.

Stochastic forcing allows for the use of data-driven models outside of the dataset from which the EOFs are obtained and the parametrization of the stochastic forcing ultimately remains a modeling choice. In \cite{resseguier2020data} a data-driven parametrization was compared to a self-similar parametrization, using SALT in the quasi-geostrophic equations. It was found that both parametrizations accurately predict numerical errors and possess good uncertainty skills. In this paper, we extend the work presented by \cite{cotter2019numerically} of stochastic forcing for the two-dimensional Euler equations.  The extension presented in this work consists of the inclusion of additional information in the data-driven approach. This information is readily available from the EOF procedure and is used to define two additional types of stochastic processes. Providing a space-time array of measurements to the algorithm yields the EOFs, which are spatial profiles, and the amplitudes of the EOFs in order to reconstruct the input measurements. The amplitudes of each EOF are a time series and provide the data that are used in this paper to calibrate stochastic processes for each of the EOFs. In order to mimic the measurements, we generate signals that have the same probability distribution function as the measured time series or have similar temporal correlation. By retaining these statistical quantities in the modeled time series, the forcing stays true to characteristic features of the measurements.


The following numerical experiments and findings are reported in this paper. We perform a direct numerical simulation (DNS) of the two-dimensional Euler equations on the unit sphere, subject to slip boundary conditions. We measure the difference between trajectories of particles advected by the fully-resolved velocity field and the corresponding filtered velocity field. The EOFs and time series that represent the amplitudes of the EOFs are obtained from this data. Stochastic ensembles are generated using three stochastic processes: Gaussian noise, noise based on the underlying pdf of the EOF time series, and noise with a temporal correlation similar to that of the EOF time series. The process of developing the time series into stochastic processes is explained in detail in a subsequent section of the paper. The results presented in this paper show that using the developed stochastic processes leads to a reduction of the ensemble mean error and ensemble spread, compared to using Gaussian noise. This is further explored by performing statistical tests for ensemble solutions. The latter is done for time scales on which data may be assimilated, where the numerical SPDE results may serve as input \cite{cotter2020particle}.

The paper is structured as follows. In section \ref{subsec:governingeqns} we introduce the deterministic and stochastic governing equations and describe the numerical experiment in detail. This is followed by a description of the data acquisition procedure in section \ref{subsec:dataacquisition}. The method used for generating random signals as a model for the measured data is described in section \ref{subsec:randomsignals}. The results of the numerical experiments are presented in section \ref{sec:ensembleforecast}. In section \ref{subsec:predictionhorizon} a maximal prediction horizon is established and in section \ref{subsec:definition_reference} an adapted reference solution defined. These results aid the uncertainty quantification of ensemble predictions, presented in \ref{subsec:uq}. Predictions on much shorter timescales are further assessed in section \ref{subsec:statisticaltests}, comparing ensemble statistics, rank histograms and conditional distributions. We conclude the paper in section \ref{sec:conclusions} and specify future challenges.


\section{\label{sec:eulereqns}SPDE formulation and stochastic models}
This section presents the formulation of the stochastic Euler equations using the SALT approach (Subsection \ref{subsec:governingeqns}), the data acquisition procedure (Subsection \ref{subsec:dataacquisition}) and the derivation of the stochastic models (Subsection \ref{subsec:randomsignals}).

\subsection{\label{subsec:governingeqns}Governing equations and flow conditions}
The two-dimensional Euler equations are central to this work. These equations are determined fully by the evolution of the vorticity dynamics \cite{zeitlin2018geophysical}. The behaviour of the vorticity $\omega$ in terms of the velocity $\mathbf{u}$ and streamfunction $\psi$ is given by \begin{align}
\partial_t \omega + (\mathbf{u}\cdot\nabla)\omega & = Q-r\omega, \label{eq:advectioneqn}\\
\mathbf{u}&=\nabla^\perp\psi, \label{eq:gradperp_psi}\\
\Delta\psi&=\omega, \label{eq:Laplaceeqn}
\end{align} which are solved on the unit square, denoted by $\mathcal D$. The perpendicular gradient $\nabla^\perp$ is defined as $(-\partial_y, \partial_x)$. A forcing and a damping term are added to the equations in order to drive the flow to a nontrivial statistically steady state. In particular, $Q(x,y) = 0.1\sin (8\pi x)$ and  $r=0.01$, which enforce eight spatial gyres that are constant in time. A slip boundary condition is applied via \begin{equation}
    \psi|_{\partial\mathcal{D}}=0
\end{equation}
along the boundary $\partial \mathcal{D}$ of $\mathcal{D}$. For this system a characteristic time scale is the large eddy turnover time, here estimated to be 2.5 time units \cite{cotter2019numerically}.


The stochastic equations associated with the Euler equations follow from the principle of stochastic advection by Lie transport (SALT) for ideal fluid dynamics \cite{holm2015variational}. In this approach, SPDEs are derived from a variational principle. In fact, a stochastically constrained functional is minimised to obtain an SPDE which retains the geometric properties equivalent to the corresponding PDE. The result is that quantities that are advected along an infinitesimal vector field $\mathbf{u}\text{d}t$ in the deterministic setting are advected along an infinitesimal vector field $\bar{\mathbf{u}}\text{d}t+\sum_i\boldsymbol\xi_i \circ \text{d}B_t^i$ in the stochastic setting. In this paper, $\bar{\cdot}$ denotes a filtered field representative of scales that can be resolved accurately on a coarse numerical grid. As a rough rule of thumb, the resolved scales would comprise of structures for which $\Delta \gtrsim mh$ where $h$ denotes the uniform grid spacing and $m$ is a factor that quantifies the desired accuracy requirements. Typically, one may think of $m\gtrsim 4$ for second order accurate methods \cite{geurtsfrohlich}. The velocity fields $\boldsymbol\xi_i$ are defined as the eigenvectors of the velocity-velocity correlation tensor \cite{holm2015variational}, $B_t^i$ is a general stochastic process. The symbol $\circ$ implies that the stochastic integral should be understood in the Stratonovich sense. This means that the integral is approximated by Riemann sum defined on the midpoints of the subintervals. For a good introduction to this material \cite{kloeden1992stochastic} and \cite{higham2001algorithmic} can be consulted.

Since the velocity field $\mathbf{u}$ is divergence-free, each velocity field $\boldsymbol \xi_i$ is divergence-free \cite{cotter2019numerically} and can be expressed by a potential function $\zeta_i$ via $\boldsymbol\xi_i = \nabla^\perp \zeta_i$. The advection velocity can then be written in terms of the potential as \begin{equation}
    \bar{\mathbf{u}}(t)\text{d}t + \sum_i\boldsymbol\xi_i\circ\text{d}B_t^i = \nabla^\perp\bar{\psi}(t)\text{d}t + \sum_i\nabla^\perp\zeta_i\circ\text{d}B_t^i.
\end{equation}
In this equation the filtered variables are used since the aim of the stochastic model is to represent the components of the fine-grid solution that are not resolvable on the coarse grid.
The resulting SPDE then reads \cite{cotter2019numerically}
\begin{align}
    \text{d}\bar{\omega} &+ \nabla^\perp \left( \bar{\psi}\text{d}t+\sum_i\zeta_i\circ\text{d}B_t^i\right)\cdot\nabla \bar{\omega} = (Q-r\bar{\omega}) \text{d}t, \label{eq:EulerSALTadv}\\
    \Delta \bar{\psi} & = \bar{\omega}. \label{eq:EulerSALTPoisson}
\end{align}

\subsection{\label{subsec:dataacquisition}Data acquisition}
The numerical method for the solution of \eqref{eq:EulerSALTadv}-\eqref{eq:EulerSALTPoisson} and the flow parameters are the same as those used in earlier studies \cite{cotter2019numerically, cotter2020particle}.  A full description of the numerical implementation can be found in the former references. Here, for completeness, we illustrate the key aspects. A finite element method is employed to solve the system of equations \eqref{eq:EulerSALTadv} and \eqref{eq:EulerSALTPoisson}. The Poisson equation for the streamfunction is discretized using a continuous Galerkin scheme. The vorticity equation \eqref{eq:advectioneqn}, including the stochastic terms, is discretized using a discontinuous Galerkin scheme. The space of discontinuous test functions guarantees numerical conservation of energy in the absence of source terms \cite{bernsen2006dis}.

Numerical time integration is performed by applying a third-order strong stability preserving Runge-Kutta (SSPRK3) method \cite{shu1988efficient}. Writing the stochastic advection equation \eqref{eq:EulerSALTadv} in the general Stratonovich SPDE form
\begin{equation}
    \text{d}\bar{\omega} = L(\bar{\omega})\text{d}t + \sum_{i=1}^m G^i(\bar{\omega})\circ\text{d}B_t^i, \label{eq:StratSPDE}
\end{equation}
where \begin{equation}
\begin{aligned}
    L(\bar{\omega}) &= -\nabla^\perp\bar{\psi}\cdot \nabla \bar{\omega} + (Q-r\bar{\omega}), \\
    G^i(\bar{\omega}) &= -\nabla^\perp\zeta_i \cdot\nabla \bar{\omega},
    \end{aligned}
\end{equation}
the SPDE \eqref{eq:StratSPDE} is integrated in time via
\begin{equation}
    \begin{aligned}
        \bar{\omega}_{(1)} &= \bar{\omega}_n + \Delta t L(\bar{\omega}_n) + \sum_{i=1}^m G^i(\bar{\omega}_n)\Delta B^i_n, \\
        \bar{\omega}_{(2)} & = \frac{3}{4}\bar{\omega}_{n} + \frac{1}{4}\left[ \bar{\omega}_{(1)} + \Delta t L\left(\bar{\omega}_{(1)}\right) + \sum_{i=1}^n G^i\left(\bar{\omega}_{(1)}\right)\Delta B^i_n \right], \\
        \bar{\omega}_{n+1} & = \frac{1}{3}\bar{\omega}_{n} + \frac{2}{3}\left[ \bar{\omega}_{(2)} + \Delta t L\left( \bar{\omega}_{(2)}\right) + \sum_{i=1}^m G^i\left(\bar{\omega}_{(2)}\right)\Delta B^i_n \right].
    \end{aligned} \label{eq:SSPRK3}
\end{equation}
The subscript $n$ denotes the $n^\mathrm{th}$ numerical time step. The stages of the Runge-Kutta algorithm are denoted by the subscripts $(1)$ and $(2)$. The time step size is given by $\Delta t$ and is chosen such that the CFL number does not exceed $1/3$. Here $\Delta B^i_n$ denote random samples drawn from an assumed probability distribution with variance $\Delta t$. For deterministic systems, the functions $G^i$ equal zero. 

The term $\nabla^\perp(\sum_i\zeta_i\circ\text{d}B_t^i)$ in \eqref{eq:EulerSALTadv} is unknown in the coarsened description and needs to be modelled. The latter is approximated as follows: \begin{equation}
    \mathbf{f}(x,t)\text{d}t := \left(\mathbf{u}(x,t)-\bar{\mathbf{u}}(x,t)\right)\text{d}t \approx \sum_{i} \boldsymbol\xi_i(x)\circ\text{d}B_t^i. \label{eq:modelgoal}
\end{equation}
The left hand side of \eqref{eq:modelgoal} accounts for the small-scale velocity fluctuations that are not resolved by coarse numerical grids. The right hand side incorporates these fluctuations as a stochastic forcing. The fluctuations are measured from high-resolution numerical data obtained from DNS of the deterministic system of equations \eqref{eq:advectioneqn}-\eqref{eq:Laplaceeqn}. The process of measuring $\mathbf{f}(x,t)$ is elaborated below.

\begin{figure}[h!]
    \centering
    \includegraphics[width=1\textwidth]{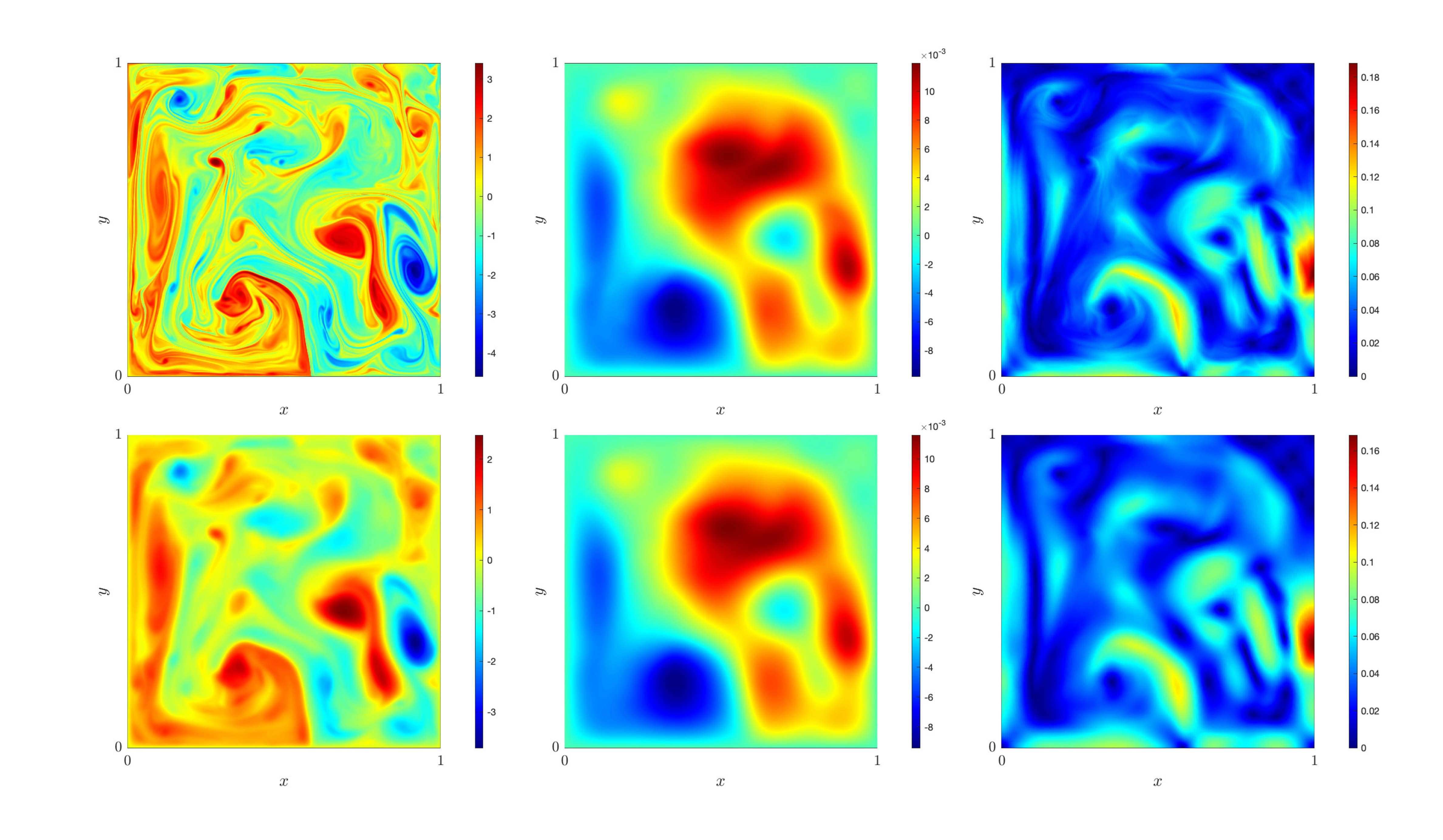}
    \caption{Fine-grid fields of the vorticity (left), streamfunction (middle) and velocity magnitude (right) after the spin-up interval. The top row shows the unfiltered fields, the bottom row shows the corresponding fields after applying the filter \eqref{eq:HelmholtzFilter}.}
    \label{fig:fields_after_spinup}
\end{figure}

A grid with $512^2$ computational cells is adopted for the DNS and all subsequent stochastic results are obtained on a coarse grid of $64^2$ computational cells. The filtered fields are derived from the fine-grid DNS results and are obtained by applying a Helmholtz operator to the streamfunction. Given a streamfunction $\psi$, the filtered streamfunction $\bar{\psi}$ is obtained by solving \begin{equation}
    (I-c\nabla^2)\bar{\psi}=\psi, \label{eq:HelmholtzFilter}
\end{equation}
where $c=1/64^2$ to filter out length scales smaller than the coarse grid size. The numerical resolutions and the filter width coincide with those adopted in \cite{cotter2019numerically}. The filtered vorticity $\bar{\omega}$ and filtered velocity $\bar{\mathbf{u}}$ are recovered from applying the relations \eqref{eq:gradperp_psi} and \eqref{eq:Laplaceeqn} to $\bar{\psi}$. The initial vorticity is prescribed, as \begin{equation}
    \omega_0 = \sin(8 \pi x)\sin(8 \pi y) + 0.4\cos(6\pi x)\cos(6\pi y) + 0.3\cos(10\pi x)\cos(4\pi y) + 0.02\sin(2\pi y) + 0.02\sin(2\pi x),
\end{equation}
from which the system will be spun-up during an interval of 100 time units so that a statistical equilibrium is reached. The time at which this is reached is denoted by $t=0$ and the data measurements start take place at this point in time. The initial fields and corresponding filtered fields at the end of the spin-up interval are found in Fig. \ref{fig:fields_after_spinup}.

A space-time sequence of measurements for determining $\mathbf{f}$ from \eqref{eq:modelgoal} is obtained by computing the difference of Lagrangian trajectories of particles advected by the velocity field $\mathbf{u}$ and those advected by the filtered velocity field $\bar{\mathbf{u}}$. The difference is measured over a single coarse-grid time step. The particles are released on the coarse grid points and thus a difference in traveled distance can be related to each grid point. A velocity correction field is subsequently obtained by dividing the difference in trajectories by the coarse-grid time step, in a manner analogous to particle image velocimetry measurement techniques in experimental fluid flow analysis \cite{adrian2011particle}. By doing so at each measuring instance, an array $\mathbf{f}(x,t)$ of velocity fields is constructed. This space-time array of measurements is decomposed into empirical orthogonal functions (EOFs or EOF modes) \cite{lumley1967structure,hannachi2007empirical}. Here, a total of 4096 EOFs are available ($64^2$ degrees of freedom), of which the first 225 are used. These EOFs account for 90\% of the energy of the measurements. Application of the EOF algorithm to a flow that has a definite statistically steady state yields \begin{equation}
    \mathbf{f}(x,t) = \boldsymbol\xi_0(x) + \sum_{i=1}^N a_i(t)\boldsymbol\xi_i(x), \label{eq:eofreconstruction}
\end{equation}
where $\boldsymbol\xi_0(x)$ is the time-mean of the measurements, $\boldsymbol \xi_i(x)$ are the spatial EOF modes, also referred to as `topos', and $a_i(t)$ are the corresponding coefficients with reference to the measurements, also referred to as `chronos'. These are recorded as time series. The EOF modes are orthonormal with respect to the inner product, thus $\left( \boldsymbol\xi_i, \boldsymbol\xi_j \right) = \delta_{ij}$, where 
\begin{equation}
\left( \boldsymbol f, \boldsymbol g \right)=\int_\Omega\! \boldsymbol f(x)\boldsymbol g(x)\,\text{d}x
\end{equation} 
with $\Omega$ the flow domain. Due to the orthonormality, the coefficients $a_i(t)$ are readily obtained by projecting the measured velocity fields onto the basis of EOFs by \begin{equation}
    a_i(t) = \left( \mathbf{f}(x,t)-\xi_0(x), \xi_i(x) \right). \label{eq:temporalcoeff}
\end{equation}
In order to have a self-contained model which allows to obtain predictions, e.g., beyond the time span of the dataset, or as surrogate statistical sample of the flow, the time traces $a_i(t)$ will be modelled with independent stochastic processes. This will be described in the next section, where also the possible connection to the available data will be elaborated.

\subsection{\label{subsec:randomsignals}Generating random signals}
We will now introduce the models for the time traces \eqref{eq:temporalcoeff} and subsequently describe how random signals are generated using these models. By comparing \eqref{eq:eofreconstruction} with \eqref{eq:modelgoal} it is clear that modelling $B_t^i(t)$ amounts to modelling $a_i(t)$. 
The following models are employed:
\begin{enumerate}
    \item The stochastic process $B_t^i$ in \eqref{eq:modelgoal} is modelled by Gaussian noise. For its discrete increments $\Delta B^i$ in \eqref{eq:SSPRK3} we use $\Delta B^i \sim \mathcal{N}(0,\Delta t)$ \cite{higham2001algorithmic}.
    \item The probability density function (pdf) of $a_i(t)$ in \eqref{eq:eofreconstruction} is estimated from the measured signals \eqref{eq:temporalcoeff} and is subsequently used to draw uncorrelated samples to compute $\Delta B^i$ in \eqref{eq:SSPRK3}.
    \item The time series $a_i(t)$ in \eqref{eq:eofreconstruction} is approximated by an Ornstein-Uhlenbeck (OU) process, using the correlation time obtained from the measurements \eqref{eq:temporalcoeff}. The constructed OU process is then used to compute $\Delta B^i$ in \eqref{eq:SSPRK3}.
\end{enumerate}


\noindent The probability distributions of model 2 are estimated by fitting a histogram to the values of the corresponding time series, yielding a separate distribution for each EOF. The histograms are fully determined by the smallest and largest measurements and the number of measurements. The number of bins is chosen as the smallest integer larger than $\sqrt[3]{2N_M}$, where $N_M$ denotes the number of measurements, i.e., the length of the time series. This choice minimizes the asymptotic mean squared error of the histogram as an estimator of the underlying pdf \cite{wilks2011statistical}. Uncorrelated samples from these distributions are drawn using inverse transform sampling. In the latter a random number $x$ is drawn from a uniform distribution between 0 and 1, which can intuitively be thought of as a probability of an event happening, and subsequently the largest value $X$ is found such that $P(X\leq x)$ holds for the estimated distribution \cite{devroye2006nonuniform}.

In model 3, the noise generated using the OU process mimics the temporal correlation of the measured time series. Denoting by $B_t^i$ the approximation of the time series $a_i(t)$, the OU process is defined as \cite{pope2001}\begin{equation}
    \text{d}B^i_t=-B^i_t\frac{\text{d}t}{T_i}+\left(\frac{2\sigma_i^2}{T_i}\right)^{1/2}\text{d}W_t^i,
\end{equation}
where $\text{d}W$ are Wiener increments and we set $T_i$ and $\sigma_i$ to be the correlation time and the standard deviation of the measured time series. These variables are determined for each EOF separately. Here, the correlation time is defined as the smallest time at which the autocorrelation function of the time series is smaller than the computed $95\%$ confidence bound. 

A consistent choice for a fourth model is one that incorporates the measured temporal correlation, whilst retaining the estimated probability distribution of measurements. However, for this approach no tractable algorithm to generate the stochastic processes was found.


In the next section, we assess the proposed stochastic models by comparing simulations on the SPDE models to findings from deterministic reference solutions.

\section{\label{sec:ensembleforecast}Assessment of forecast ensembles}
In this section, we provide results of forecast ensembles using the aforementioned methods to generate stochastic signals that serve to force the coarsened dynamics. We first identify a maximal prediction horizon for assessing the forecast ensembles. An adapted reference solution is defined based on the measurements, incorporating on the coarse numerical grid the measured effects of small-scale motions. Subsequently, we show results of forecast ensembles. Statistics are computed and compared to the filtered DNS and the adapted reference solution to quantitatively compare the different stochastic forcing methods.

\subsection{\label{subsec:predictionhorizon}Establishing a maximal prediction horizon}
In order to define the maximal prediction horizon until which stochastically forced coarse numerical solutions can reasonably be compared to the DNS results, we set up the following numerical experiment. Starting from an initial condition on the fine grid, we generate a set of perturbed initial conditions of which we then follow the evolution over time. The perturbations are applied in Fourier space by shifting the phase of the Fourier coefficients, while keeping the amplitudes the same. The phase shift is applied only to modes of wave lengths smaller than the smallest scale resolved by the corresponding coarse grid. That is, only unresolved scales of the coarse grid are perturbed, leaving the resolved modes unaltered. Specifically, a value $l$ is chosen and all Fourier modes with wave numbers $|k|=(k_x^2 + k_y^2)^{1/2} \in [l, l+1)$ are affected by the additional phase shift. Here $k_x$ and $k_y$ denote the wave numbers in the $x-$ and $y-$direction, respectively, and $l$ is chosen as $64, 128$ and $256$. The phase shift is set to $\pi$ to satisfy the boundary conditions.

As time evolves, the initial perturbation increasingly affects the resolved scales, up to the point where the instantaneous resolved fields will be entirely different from each other. We define this point of no longer truthfully following the unperturbed solution as the maximal prediction horizon $T_\mathrm{max}$, after which no model can be expected to consistently give accurate point-wise predictions owing to the sensitivity of the evolving solution to the initial conditions. The value of $T_\mathrm{max}$ is expected to depend on the choice of perturbed modes and choice of simulation parameters. However, in this numerical experiment it serves to provide an estimate of the maximal predication horizon.

\begin{figure}[h!]
    \centering
    \begin{subfigure}[t]{0.45\textwidth}
    \includegraphics[width=\textwidth]{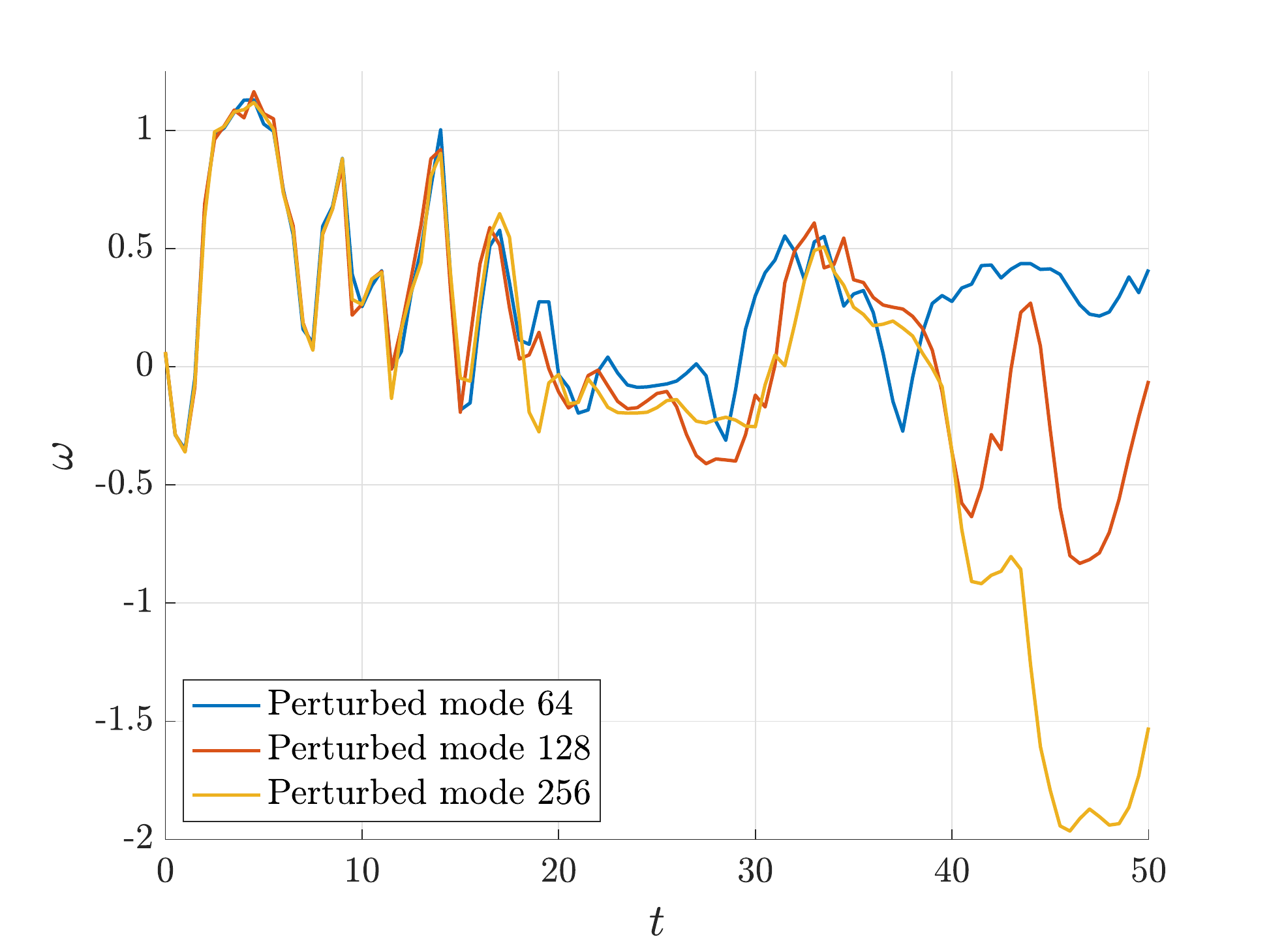}
    \end{subfigure}
    \begin{subfigure}[t]{0.45\textwidth}
    \centering
    \includegraphics[width=\textwidth]{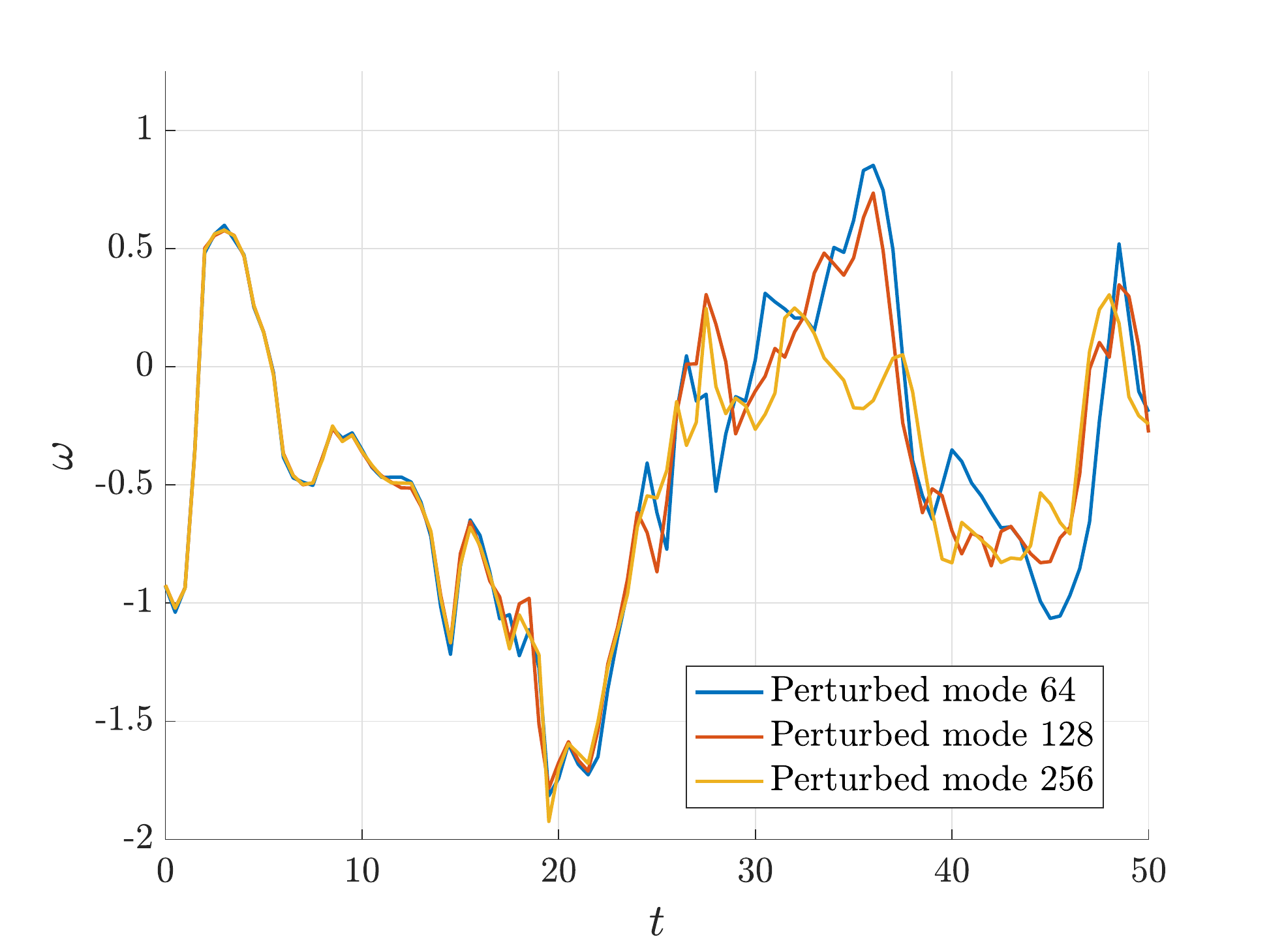}
    \end{subfigure}
    \caption{Development of the vorticity in two points of the domain, obtained by DNS of perturbed fine-grid initial conditions. On the left, the vorticity is shown at $(0.25,0.25)$ and on the right at $(0.75,0.75)$. Each initial condition is obtained by phase-shifting small-scale modes of the streamfunction field. In the results shown here, the Fourier modes with wave numbers $|k|=(k_x^2 + k_y^2)^{1/2}\in [l, l+1) $, $l = 64, 128, 256$ are phase-shifted by $\pi$.}
    \label{fig:PerturbedDNS}
\end{figure}

The observed behaviour following the small-scale phase-shift perturbations is illustrated in Fig. \ref{fig:PerturbedDNS}. The evolution of the vorticity using the perturbed initial conditions has been measured on four illustrative points in the domain, at $(0.25, 0.25), (0.25,0.75), (0.75,0.25)$ and $(0.75,0.75)$, of which two points are shown in the figure. It can be seen that the evolution of the vorticity values at the measured points in the domain is initially indistinguishable. At $t=10$ slight differences are visible and at $t=20$ the measured values are markedly different. The latter result is especially clear at the point $(0.25,0.25)$, in the left figure. Thus, we conclude that subsequent stochastic realizations can not be reasonably assessed after $t=20$, which we set as the value for the maximal prediction time $T_\mathrm{max}$. 

\subsection{Defining the reference solution} \label{subsec:definition_reference}
In order to compare the different stochastic models one has to define a reference solution. The choice of the latter is not unique. In this work we define two reference solutions that are employed to measure performance of a given forcing model. The first one is the filtered fine-grid solution, employing the filter \eqref{eq:HelmholtzFilter}, and is indicative of flow scales that can be resolved on the coarse grid. Next to the filtered fine-grid solution, we define a reference solution as the numerical solution of \eqref{eq:EulerSALTadv}-\eqref{eq:EulerSALTPoisson} where the reconstructed signal \eqref{eq:eofreconstruction}, \eqref{eq:temporalcoeff} is used in \eqref{eq:modelgoal} instead of the stochastic forcing. This provides a prescribed deterministic forcing for the coarse numerical simulation. We call this the adapted reference solution. We note that the structure of the closure term \eqref{eq:modelgoal} does not account for discretization error and is itself not an exact closure since the noise is introduced only in the advection velocity. The inclusion of discretization error is what sets the filtered DNS and the adapted reference solution apart. Therefore, by comparing the stochastic ensembles against the adapted reference solution, one is able to distinguish between modelling error from the proposed stochastic models and the discretization error. 

\begin{figure}[h!]
    \centering
    \includegraphics[width=\textwidth]{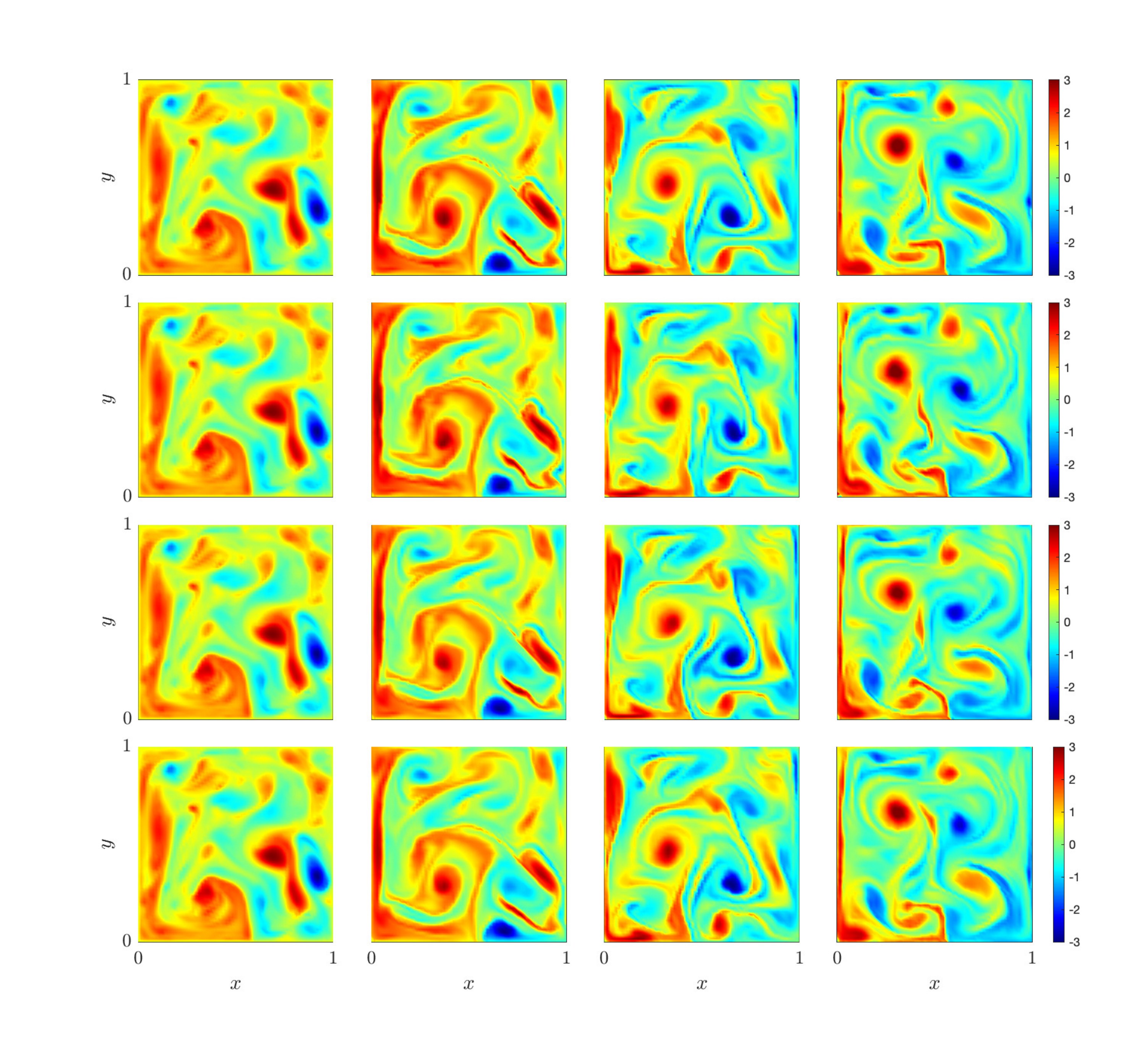}
    \caption{Coarse-grid fields of the vorticity at various points in time. The top row shows, from left to right, the adapted reference solution at $t=0, t=10, t=20$ and $t=30$. The other rows show realizations of stochastically forced numerical solutions at the aforementioned times. The second row uses Gaussian noise, the third row uses random samples from estimated distributions and the bottom row uses OU processes.} 
    \label{fig:plot_array}
\end{figure}

The adapted reference solution at $t=0, 10, 20, 30$ is shown in the top row of Fig. \ref{fig:plot_array}. At the same points in time, a single realization of each of the stochastically forced solutions is shown. The second row shows a realization using Gaussian noise, the third row using estimated pdfs and the bottom row using OU processes. While slight differences between the various realizations can be observed, the qualitative behaviour seems indistinguishable. A more detailed, quantitative comparison of the methods is provided in the following subsections.

\subsection{\label{subsec:uq}Uncertainty quantification of ensemble predictions}

The evolution of the vorticity and streamfunction is used for uncertainty quantification. First, the ensemble predictions are compared globally to the reference solution. In this subsection, the ensembles are compared only the adapted reference solution so that accumulation of discretization error in the coarse numerical solutions is not included in the comparison. Subsequently, similar to \cite{cotter2019numerically} four points in the domain are picked for pointwise uncertainty quantification. For each point one ensemble standard deviation around the ensemble mean solution is shown and compared to the reference solution at the same point.  In these tests, the ensemble is initialized from a single initial condition in order to isolate the effects of the stochastic processes on the uncertainty of the numerical solution. The initial condition is obtained by injecting the DNS vorticity field onto the coarse grid. Each SPDE is simulated up to $T_\mathrm{max}=20$, and every ensemble is composed of 200 realizations of the SPDE. Our interest here lies in comparing the errors and spreads for the different types of stochastic processes used in the forcing \eqref{eq:modelgoal}. Different error measures will be monitored as outlined next.

For global comparison to the reference solution, we define the quantity \begin{equation}
    \frac{(\omega(x,t), \omega_{\mathrm{ref}}(x,t))}{(\omega_\mathrm{ref}(x,t), \omega_\mathrm{ref}(x,t))} = \frac{\int\! \omega(x,t)\omega_\mathrm{ref}(x,t)\,\text{d}x}{\int\! \omega_\mathrm{ref}(x,t)\omega_\mathrm{ref}(x,t)\,\text{d}x},
    \label{eq:globallikenessmeasure}
\end{equation}
which can be considered a measure of correlation between the vorticity $\omega$ obtained from the stochastically forced numerical solution and the vorticity $\omega_\mathrm{ref}$ obtained from the reference solution. The same quantity is computed for the streamfunction. The pointwise comparisons are acquired by measuring the instantaneous vorticity and streamfunction at several grid points.

The stochastic ensembles are assessed using the ensemble mean, ensemble standard deviation and ensemble mean error. Here, we denote an ensemble of $N$ stochastic realizations by $\{X_{i,j}\}$, where $i=1,\ldots,N$ denotes the realization and $j=0,\ldots, T$ denotes the time index. Then, the ensemble mean at time instance $j$ is defined as \begin{equation}
    \langle{X}_j\rangle = \frac{1}{N}\sum_{i=1}^N X_{i,j}, \label{eq:Measure-Mean}
\end{equation} 
and the standard deviation, here referred to as spread, is defined as \begin{equation}
    \mathrm{Spread}(X_{i,j}) = \sqrt{\frac{1}{N} \sum_{i=1}^N \left(X_{i,j}-\langle{X}_j\rangle\right)^2}. \label{eq:Measure-Spread}
\end{equation}
A small spread indicates a sharp ensemble forecast and a large spread suggests an increased uncertainty in the forecast. The reference solution $Y_j, j=0,\ldots, T$ is computed at the same time instances as $\{X_{i,j}\}$. The ensemble mean error of $\{X_{i,j}\}$ is then defined as \begin{equation}
\mathrm{ME}(X_{i,j} , Y_j) = \left|\langle{X}_j\rangle - Y_j\right| \label{eq:Measure-ME}
\end{equation}
A small ensemble mean error indicates that the ensemble closely follows the reference solution, whereas a large value implies that the ensemble and the reference solution have deviated considerably from each other.

\begin{figure}[h!]
    \centering
    \includegraphics[width=\textwidth]{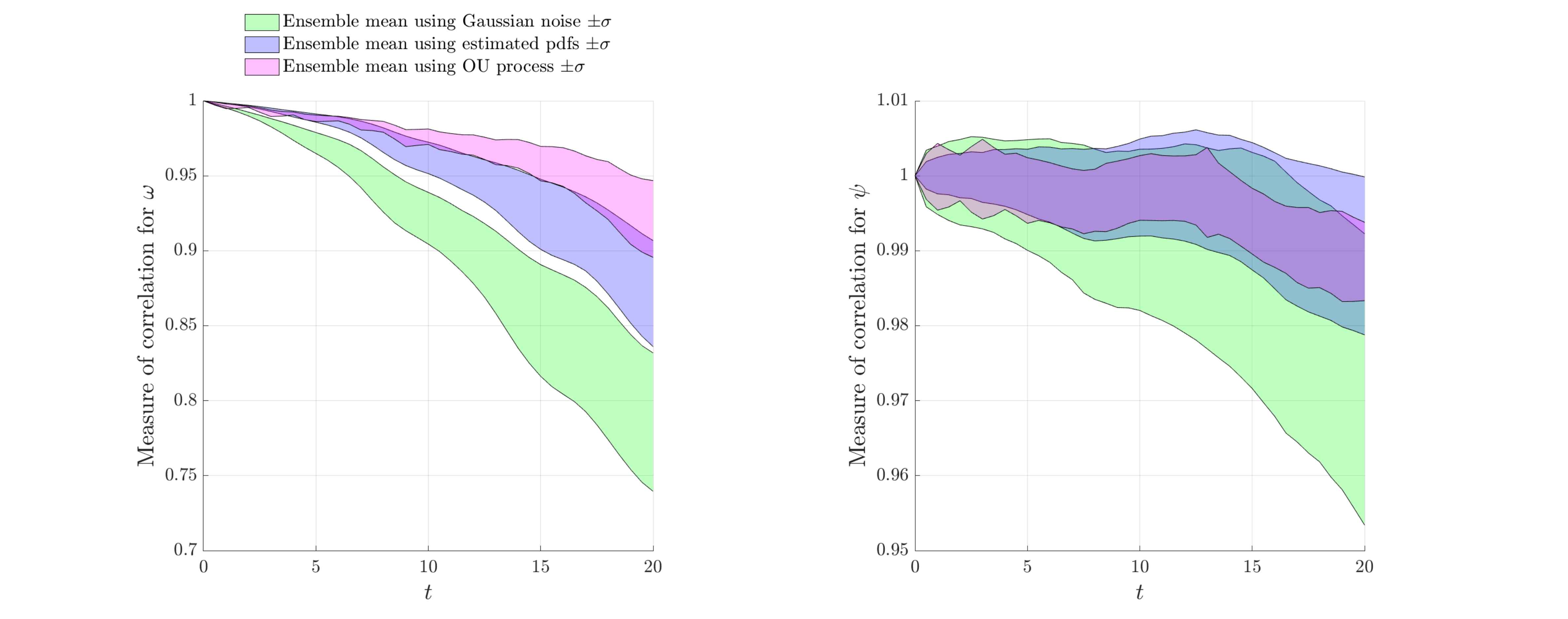}
    \caption{Measure of correlation \eqref{eq:globallikenessmeasure} between the forecast ensembles and the adapted reference solution. The left figure shows the likeness of the vorticity fields, the right figure shows the likeness of the streamfunction. Each band is defined as one ensemble standard deviation around the ensemble mean ensemble. The green band is generated using Gaussian noise, the blue band uses the estimated pdfs and the purple band uses OU processes. The results for each method are generated for an ensemble of 200 realizations.}
    \label{fig:spreads_innerproducts}
\end{figure}

The correlation measure \eqref{eq:globallikenessmeasure} is shown in Fig. \ref{fig:spreads_innerproducts} for the vorticity and the streamfunction. Using estimated pdfs or OU processes show favourable results when compared to using Gaussian noise, for both quantities. A clear difference between the methods can be observed for the vorticity on the time scale of $T_\mathrm{max}$. At this point, using estimated pdfs or OU processes yields a smaller spread than using Gaussian noise, and the results of the latter show a smaller correlation with the adapted reference solution. A significant increase in the correlation can also be observed for the streamfunction. The results using estimated pdfs or OU processes, as opposed to using Gaussian noise, exhibit both a larger likeness with the reference solution as well as a smaller spread. Compared to the ensemble obtained using Gaussian noise, at $t=20$ the ensemble standard deviation of the vorticity was found to be $23\%$ and $44\%$ when using estimated pdfs and OU processes, respectively. For the streamfunction, these values were correspondingly observed to be  $46\%$ and $73\%$. Moreover, the results for the estimated pdfs and the OU processes are nearly indistinguishable before $t=5$.

\begin{figure*}[h!]
    \centering
    \includegraphics[width=\textwidth]{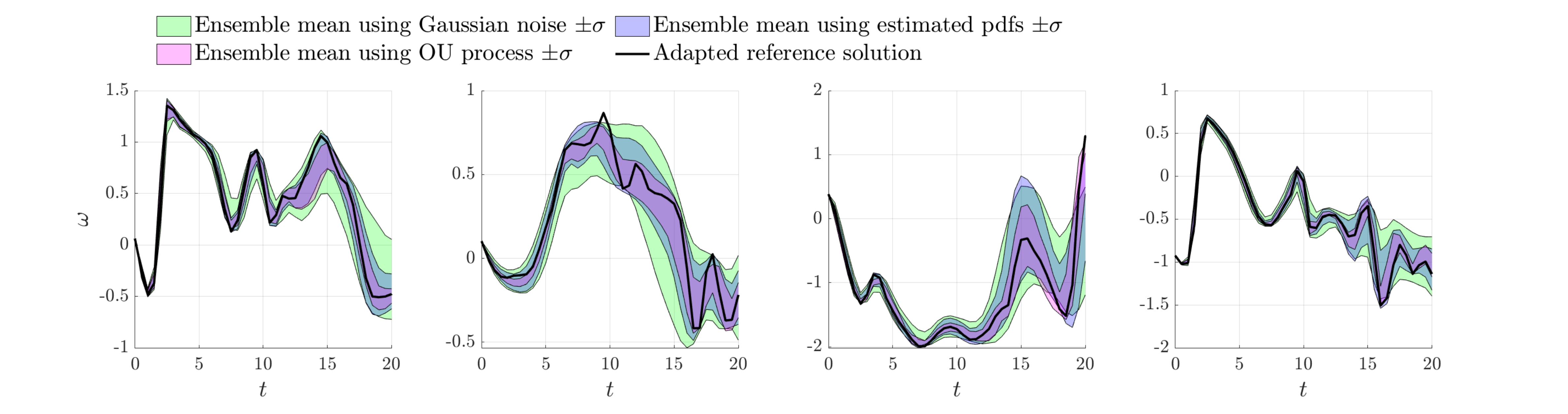}
    \caption{Vorticity measured on four points in the domain. From left to right, $(0.25,0.25), (0.25,0.75), (0.75,0.25), (0.75,0.75)$. The black lines show the development of adapted reference solution. The green band is generated using Gaussian noise, the blue band uses the estimated pdfs and the purple band uses OU processes. The results for each method are generated from an ensemble of 200 realizations.}
    \label{fig:spread_vort}
\end{figure*}

The evolution of the vorticity in four points of the domain is shown in Fig. \ref{fig:spread_vort}. The locations considered are $(0.25,0.25), (0.25,0.75), (0.75,0.25)$ and $(0.75,0.75)$. In each of these plots, the solid black line is the adapted reference solution and the colored bands present are the ensemble standard deviations around the corresponding ensemble mean. In all measured points, forcing based on Gaussian noise produces the largest spread. It is clearly visible that using the OU process yields the smallest ensemble spread and using the estimated pdfs only slightly increases the spread compared to using the OU process.

\begin{figure*}[h!]
    \centering
    \includegraphics[width=\textwidth]{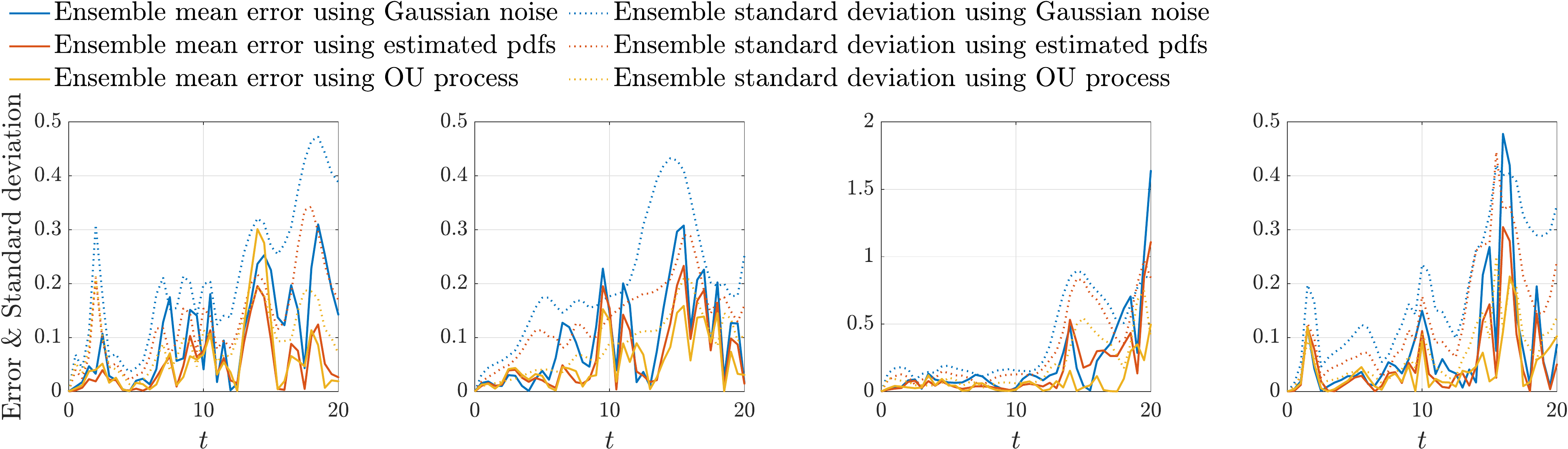}
    \caption{Ensemble mean error and ensemble standard deviation for the vorticity on four points in the domain. From left to right, $(0.25,0.25), (0.25,0.75), (0.75,0.25), (0.75,0.75)$. The ensemble mean error is depicted by the solid lines, the ensemble standard deviation by the dotted lines.}
    \label{fig:spread_stats_vort}
\end{figure*}

The ensemble mean error and the ensemble standard deviation are shown in Fig. \ref{fig:spread_stats_vort}, where the ensemble mean error \eqref{eq:Measure-ME} is taken with respect to the adapted reference solution. It becomes evident that the mean error develops similarly for each ensemble. The mean errors for ensembles using the estimated pdfs and the OU process are nearly indistinguishable until $t=10$, after which some smaller differences can be observed. In contrast, using Gaussian noise results in a much larger spread.

\begin{figure*}[h!]
    \centering
    \includegraphics[width=\textwidth]{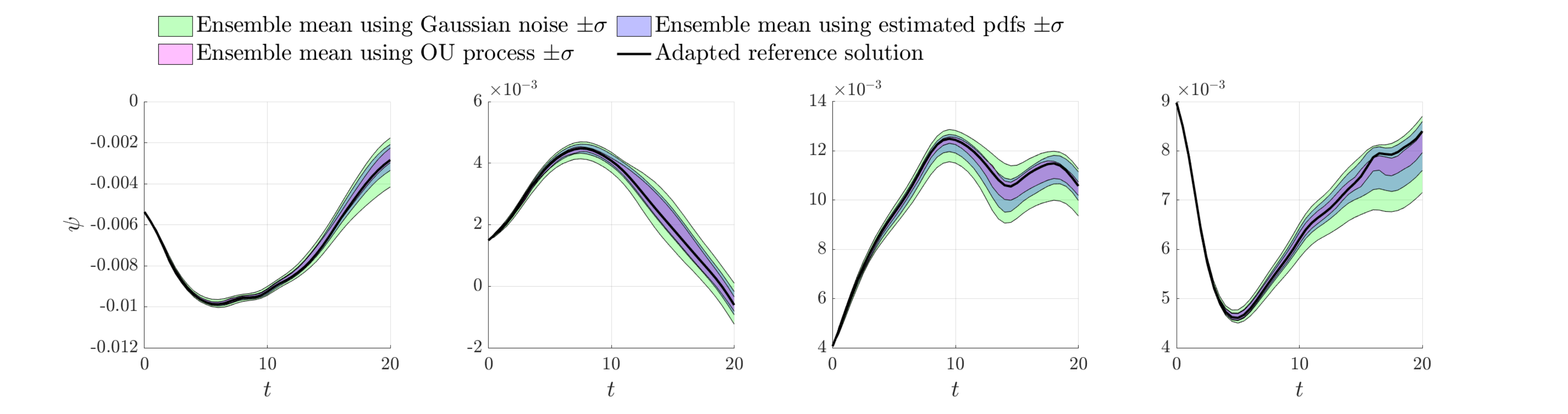}
    \caption{Streamfunction measured on four points in the domain. From left to right, $(0.25,0.25), (0.25,0.75), (0.75,0.25), (0.75,0.75)$. The black lines show the development of the adapted reference solution. The green band is generated using Gaussian noise, the blue band uses the estimated pdfs and the purple band uses OU processes. The results for each method are generated from an ensemble of 200 realizations.}
    \label{fig:spread_strm}
\end{figure*}

\begin{figure*}[h!]
    \centering
    \includegraphics[width=\textwidth]{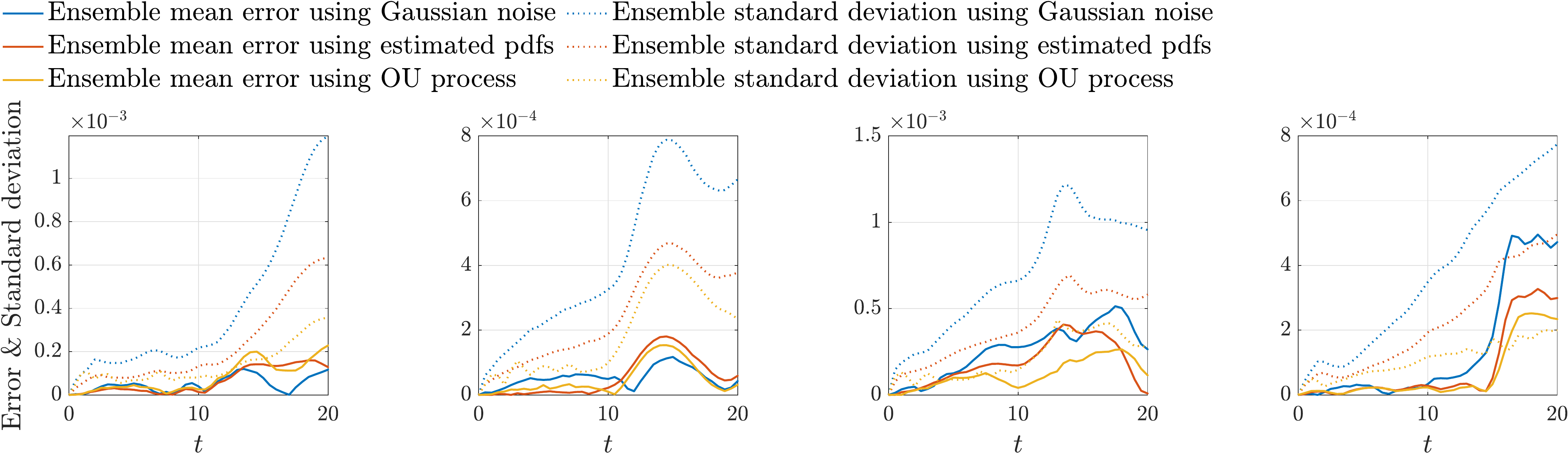}
    \caption{Ensemble mean error and ensemble standard deviation for the streamfunction on four points in the domain. From left to right, $(0.25,0.25), (0.25,0.75), (0.75,0.25), (0.75,0.75)$. The ensemble mean error is depicted by the solid lines, the ensemble standard deviation by the dotted lines.}
    \label{fig:spread_stats_strm}
\end{figure*}

Fig. \ref{fig:spread_strm} shows the development of the streamfunction in the aforementioned points of the domain. The streamfunction is a smoother function than the vorticity, which is reflected in the smooth evolution of the former. In this figure it can also be observed that all ensembles accurately capture the adapted reference solution, with the OU model performing slightly better. The plots in Fig. \ref{fig:spread_stats_strm} show the ensemble mean error and the ensemble standard deviation for the same points in the domain. Analogously to the vorticity, we find that the ensembles using the OU process and the estimated pdfs result in a smaller spread than the ensemble using Gaussian noise. Furthermore, it is observed that the ensemble mean error does not exceed the ensemble standard deviation before $t=10$ and only does so occasionally after this point in time, indicating the reference solution is captured well by the ensembles.




\newpage
In this subsection we have shown that the three considered stochastic processes accurately follow the adapted reference solution for multiple characteristic time units. Compared to Gaussian noise, using estimated pdfs or OU processes to define the stochastic forcing yielded a smaller spread of the ensemble forecast. Using a global measure, it is found that the latter two types of forcing yield ensembles that better resemble the adapted reference solution. In the next subsection, we perform additional statistical tests to assess short-time predictions.

\subsection{\label{subsec:statisticaltests}Statistical tests for ensemble forecasts}
Additional ensemble statistics are collected in order to further assess the numerical results of the SPDEs. In particular, forecast ensembles are generated for short lead times. 

Two sets of initial conditions are generated to assess the stochastic models by sampling from two reference solutions: the filtered DNS and the adapted reference solution as presented in section \ref{subsec:predictionhorizon}. The filtered DNS does not contain discretization and modelling error, whereas the adapted reference solution does.  Therefore the use of both reference methods provides insight into the effects of these errors on the statistical quantities. Two distinct sets of initial conditions are acquired by sampling the reference solutions at $t=0,5,10,\ldots,350$, measured after the spin-up time. An ensemble forecast consisting of one hundred stochastic realizations is computed for each initial condition. Every stochastic realization is run for two time units and stored every 0.04 time units in order to study the results for short lead times. This time interval is similar to time intervals at which data may be assimilated \cite{cotter2020particle}. Subsequently, the statistics are computed by comparing the ensembles to the corresponding reference solution. The statistics are provided below for both sets of initial conditions separately.

\begin{figure}[h!]
    \centering
    \begin{subfigure}[t]{0.45\textwidth}
    \includegraphics[width=\textwidth]{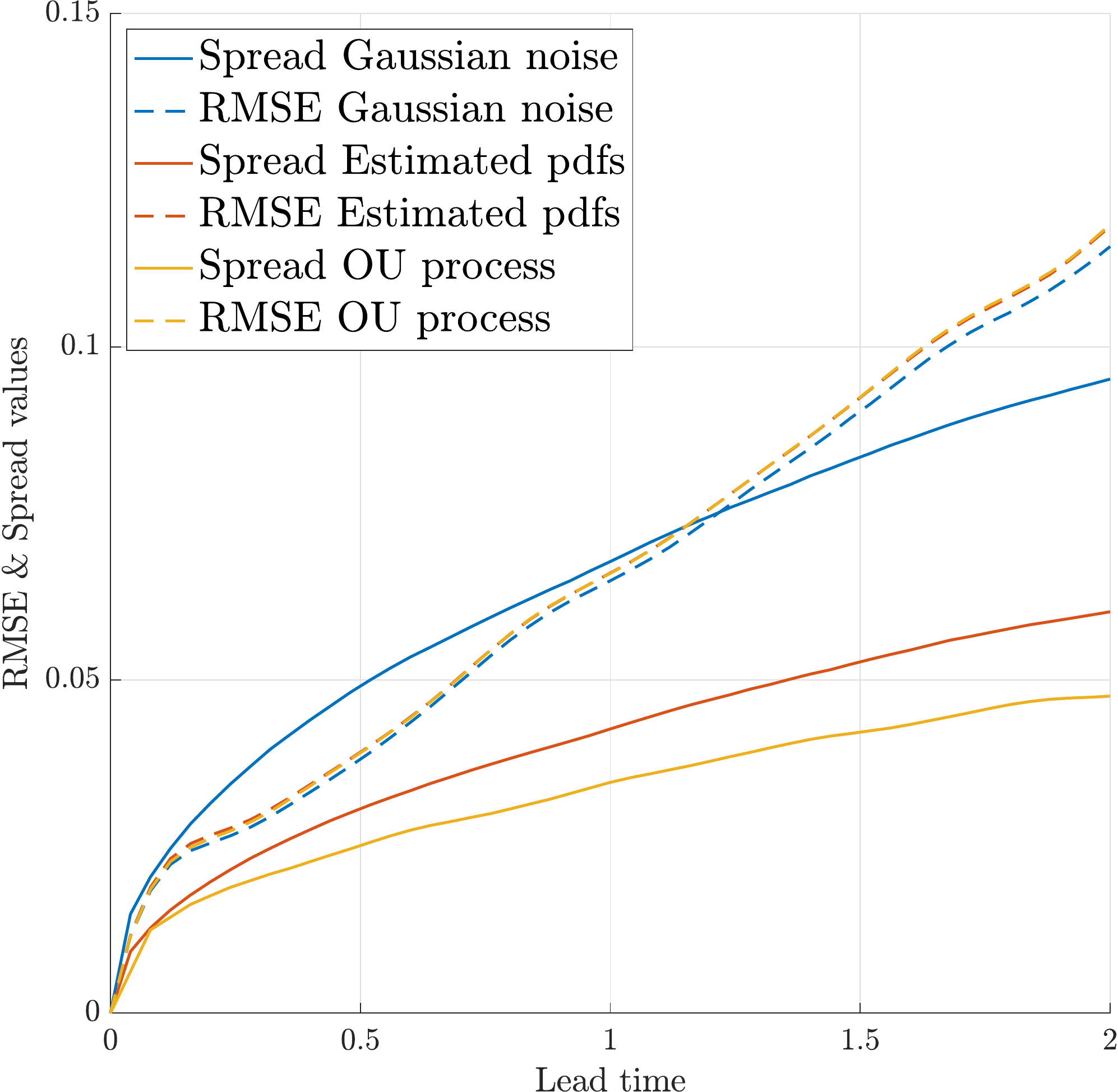}
    \end{subfigure}
    \begin{subfigure}[t]{0.45\textwidth}
    \centering
    \includegraphics[width=\textwidth]{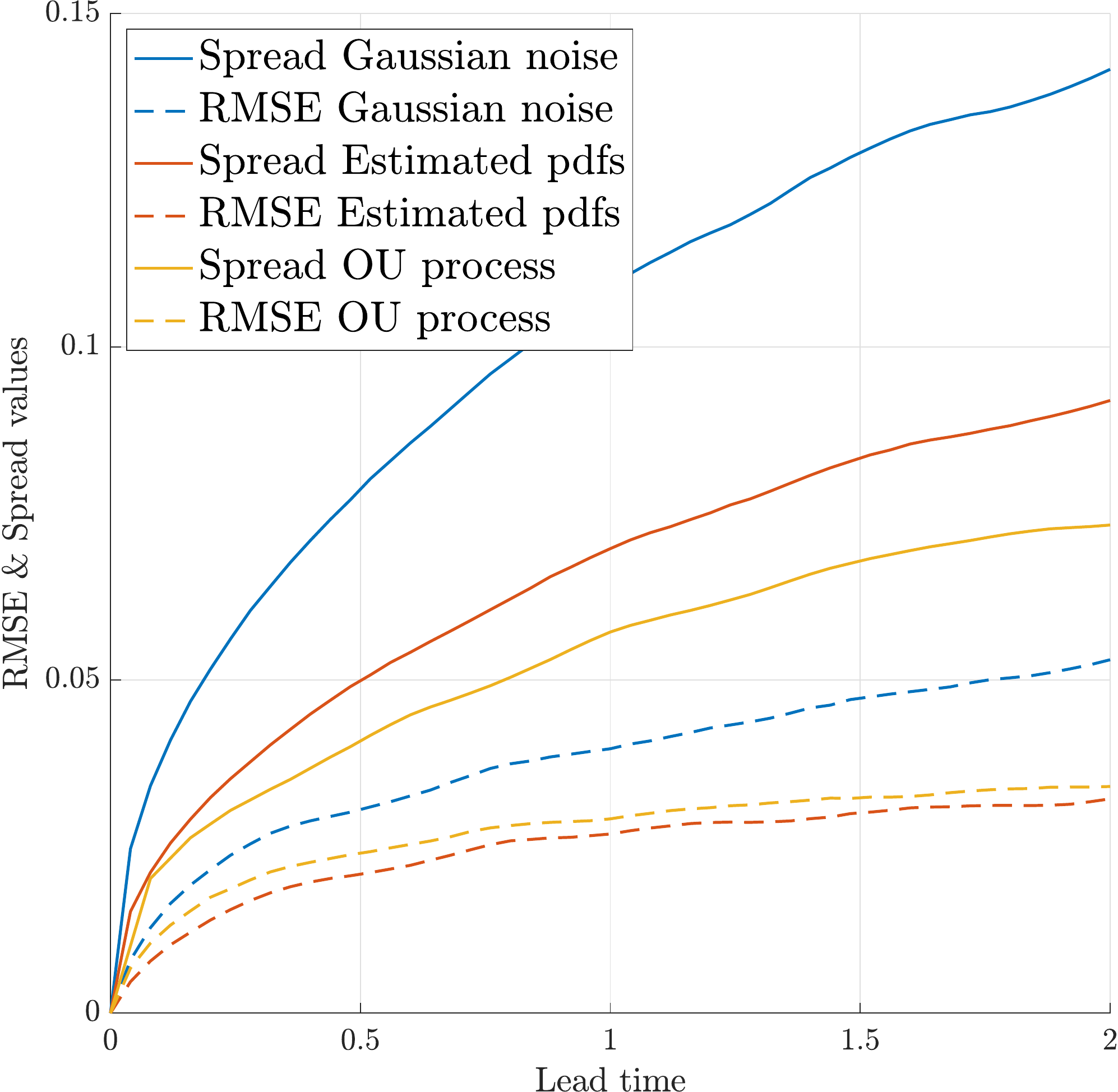}
    \end{subfigure}
    \caption{RMSE and spread as a function of time when comparing the stochastic ensembles to two different reference solutions. On the left, the filtered DNS is regarded as the reference solution and on the right the coarse simulation including the measured $\boldsymbol\xi_i$ is used. The data for each figure consists of 71 ensembles of 100 stochastic realizations each.}
    \label{fig:RMSRMEplots}
\end{figure}

As a first quantity we compute the root mean square error (RMSE). Recall that $\{X_{i,j}\}, i=1,\ldots,N, j=0,\ldots,T$ denotes an ensemble of $N$ realizations measured at $T+1$ times. The RMSE between the ensemble mean of the SPDE and the reference solution is computed from
\begin{equation}
        \mathrm{RMSE}(X_{i,j}, Y_j) =\sqrt{\frac{1}{N}\sum_{j=1}^N (\langle{X}_j\rangle-Y_j)^2}. \label{eq:Measure-RMSE}
\end{equation}
 This provides a measure for the average error of the ensemble \cite{leutbecher2009diagnosis}. The plots in Fig. \ref{fig:RMSRMEplots} show the development of the RMSE and the spread \eqref{eq:Measure-Spread} for increasing lead time for the different stochastic processes. In the left figure the stochastic ensembles are compared to the filtered DNS, in the right figure the ensembles are compared to the adapted reference solution. The RMSE values in the left graph of Fig. \ref{fig:RMSRMEplots} show rapid growth, indicating that the ensemble mean deviates quickly from the filtered DNS. In contrast, the RMSE values obtained using the adapted reference solution show a significant error reduction. This suggests that the rapid error growth in the left figure is due to the fact that the gap between the coarse-grid SPDE and the filtered DNS contains not only the modelling error but also the discretization error. In addition, the right plot in Fig. \ref{fig:RMSRMEplots} shows that using the estimated pdfs and the OU process yield similar values of the RMSE and the spreads develop comparably as well.
 
The second statistical quantity that we compute are rank histograms, which are a tool for measuring the reliability of an ensemble of forecasts \cite{hamill2001interpretation}. A rank histogram is obtained by plotting the number of occurrences of particular outcomes of the rank function. Here, the rank function $R$ keeps track of where the reference solution appears in the list of sorted ensemble members. That is, given a reference value $Y_j$ and a list of $N$ sorted ensemble members $\{X_{i,j}\}$, $R$ is equal to the integer $r$ that identifies the position of $Y_j$ in the sorted list.
It is defined as follows:
\begin{equation}
    R\left(Y_j, \{X_{i,j}\}\right) = \begin{cases} r &\text{if } Y_j \geq X_{r,j}, \\
    0 &\text{otherwise.}
    \end{cases}
\end{equation}
If the forecast is reliable, then the reference value and the stochastic realizations are indistinguishable. This means that the underlying distributions of the reference value and the stochastic realizations are the same, which implies that the reference value is equally likely to be larger than any number of ensemble members. Thus, the rank function is equally likely to take on any value between 1 and $N$ for reliable forecast ensembles and should therefore produce a rank histogram which approximates a uniform distribution. 

\begin{figure}[h!]
    \centering
    \includegraphics[width=\textwidth]{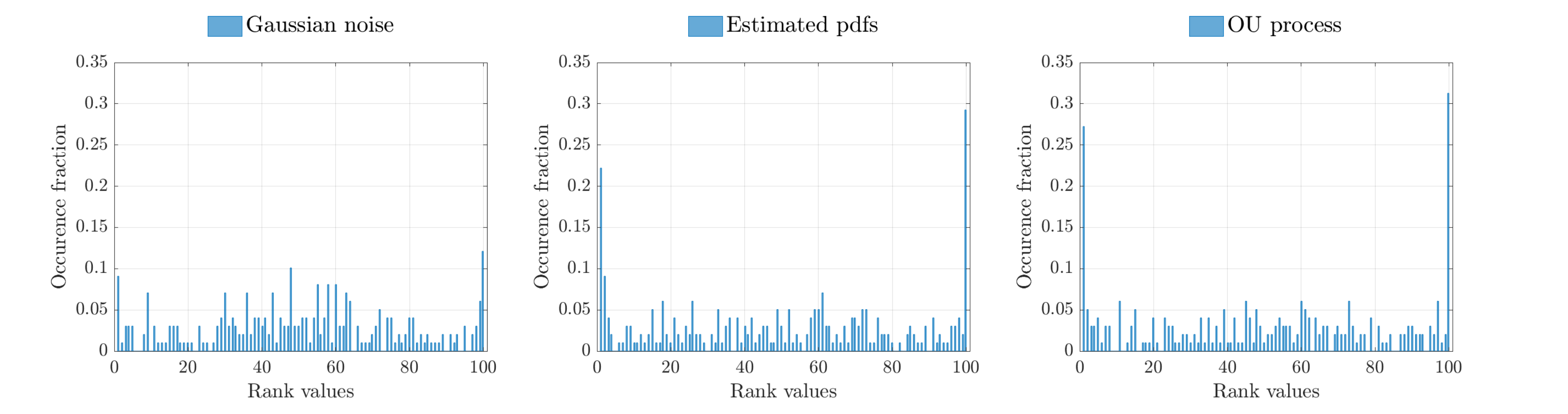}
    \caption{Rank histograms using measurements at the points $(0.25,0.25),(0.25,0.75),(0.75,0.25)$ and $(0.75,0.75)$ at a lead time of $t=0.2$. A total of 71 ensembles are computed, each consisting of 100 stochastic realizations and compared to the filtered DNS at the corresponding time.}
    \label{fig:RankHistogramsFilteredDNS}
\end{figure}

\begin{figure}[h!]
    \centering
    \includegraphics[width=\textwidth]{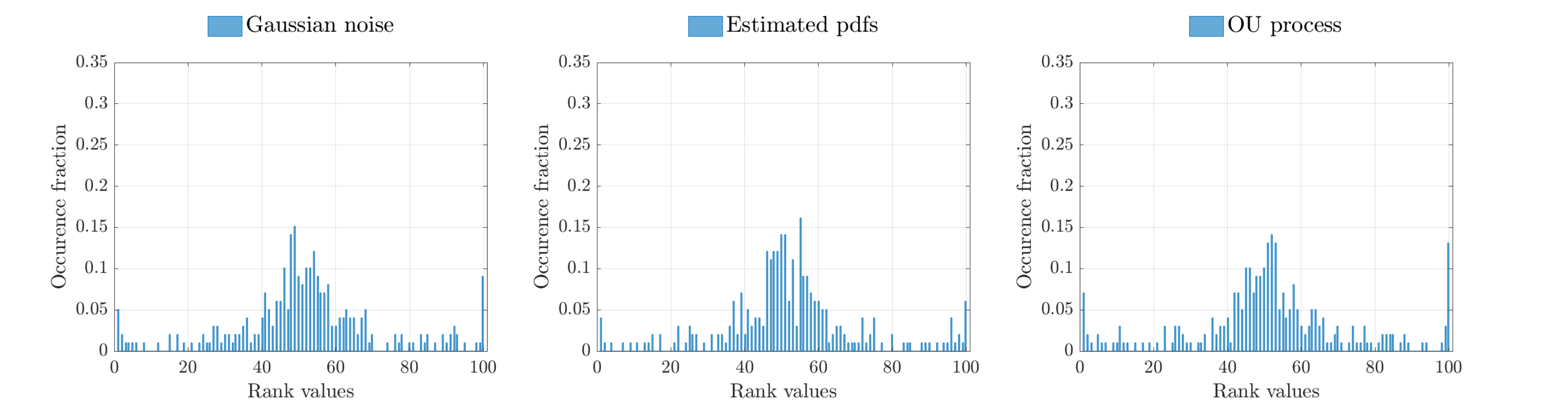}
    \caption{Rank histograms using measurements at the points $(0.25,0.25),(0.25,0.75),(0.75,0.25)$ and $(0.75,0.75)$ at a lead time of $t=0.2$. A total of 71 ensembles are computed, each consisting of 100 stochastic realizations and compared to the adapted reference solution at the corresponding time.}
    \label{fig:RankHistogramsCoarseWithXis}
\end{figure}

Figures \ref{fig:RankHistogramsFilteredDNS} and \ref{fig:RankHistogramsCoarseWithXis} show the rank histograms when using the filtered DNS and the adapted reference solution, respectively, as reference. The measurements at the points $(0.25,0.25),(0.25,0.75),(0.75,0.25)$ and $(0.75,0.75)$ at a lead time of $0.2$ time units are used to generate the histograms. Only the rank histograms at this particular lead time are shown here, rank histograms at different lead times displayed similar results.

The rank histograms using the filtered DNS (Fig. \ref{fig:RankHistogramsFilteredDNS}) show clear peaks at the edges, caused by all ensemble members either overestimating or underestimating the truth. This effect is least pronounced when applying Gaussian noise, due to the larger spread in the ensemble. The rank histograms obtained when comparing the ensembles to the adapted reference solution (Fig. \ref{fig:RankHistogramsCoarseWithXis}) show peaks around the center of the ensembles are a sign of overdispersion, indicating that the reference solution ranks within the middle range of the ensembles. This is a direct result of the small mean error. The peaks at the edges are significantly reduced when using the adapted reference solution. This is especially clear when using estimated pdfs, which indicates that these ensembles, while showing a small spread, more accurately capture the reference solution. Overall, the differences between the rank histograms of the different methods are small. This indicates that reliability of the ensembles does not seem to depend on the choice of stochastic forcing.


The third statistical quantity that is presented here is the evolution of the vorticity over different time spans, conditioned on the vorticity value at a reference time. That is, the conditional probability distribution
\begin{equation}
    P\left[\omega(t+\tau)-\omega(t)| \omega(t) = \omega_\mathrm{ref}\right] \label{eq:conditional_probability}
\end{equation}
is estimated for different values of $\tau$. This quantity describes the statistical evolution of the vorticity over a time interval of length $\tau$, given a fixed initial configuration.

\begin{figure}[h!]
    \centering
    \includegraphics[width=\textwidth]{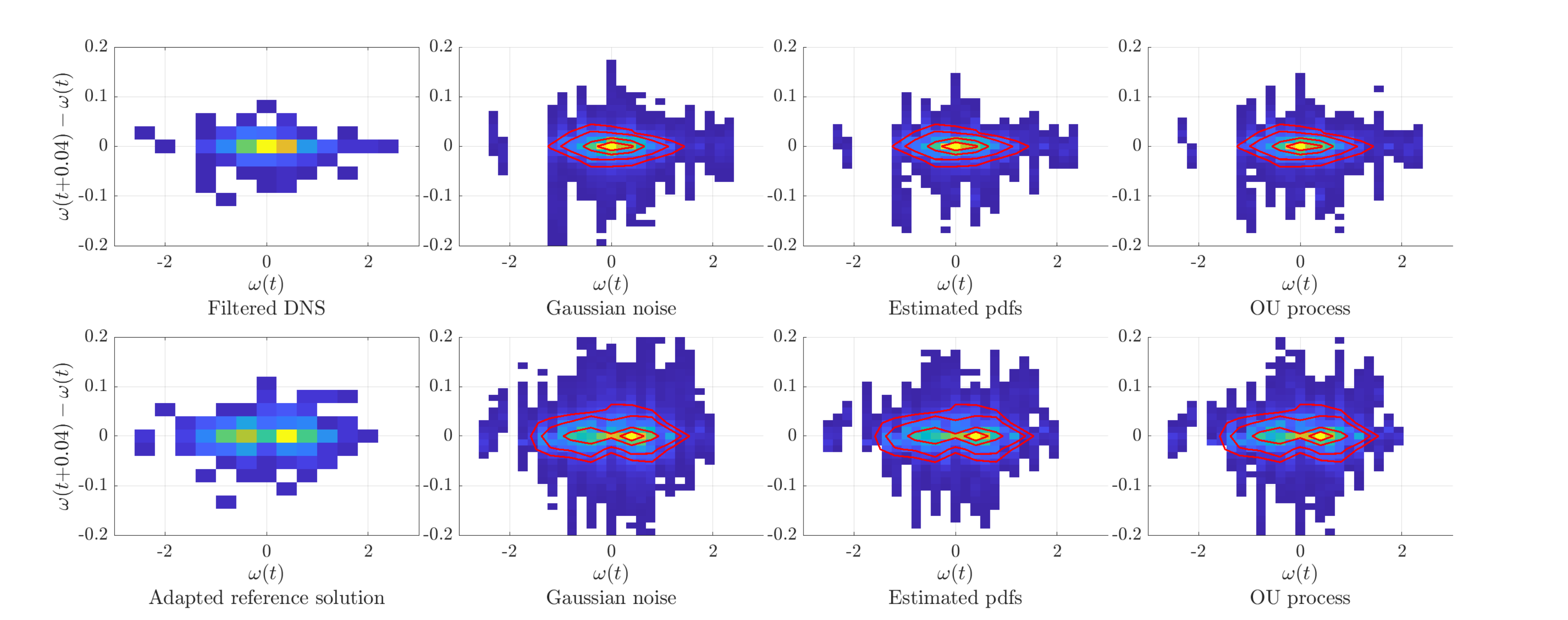}
    \caption{Conditional probability \eqref{eq:conditional_probability} for lead time $\tau=0.04$. The top row shows the distributions using the filtered DNS as a reference, the bottom row uses the adapted reference solution. The contour lines of the reference conditional distributions are overlaid on the distributions obtained from the stochastic ensembles for easier qualitative comparison.}
    \label{fig:condpdf_tau1}
\end{figure}

\begin{figure}[h!]
    \centering
    \includegraphics[width=\textwidth]{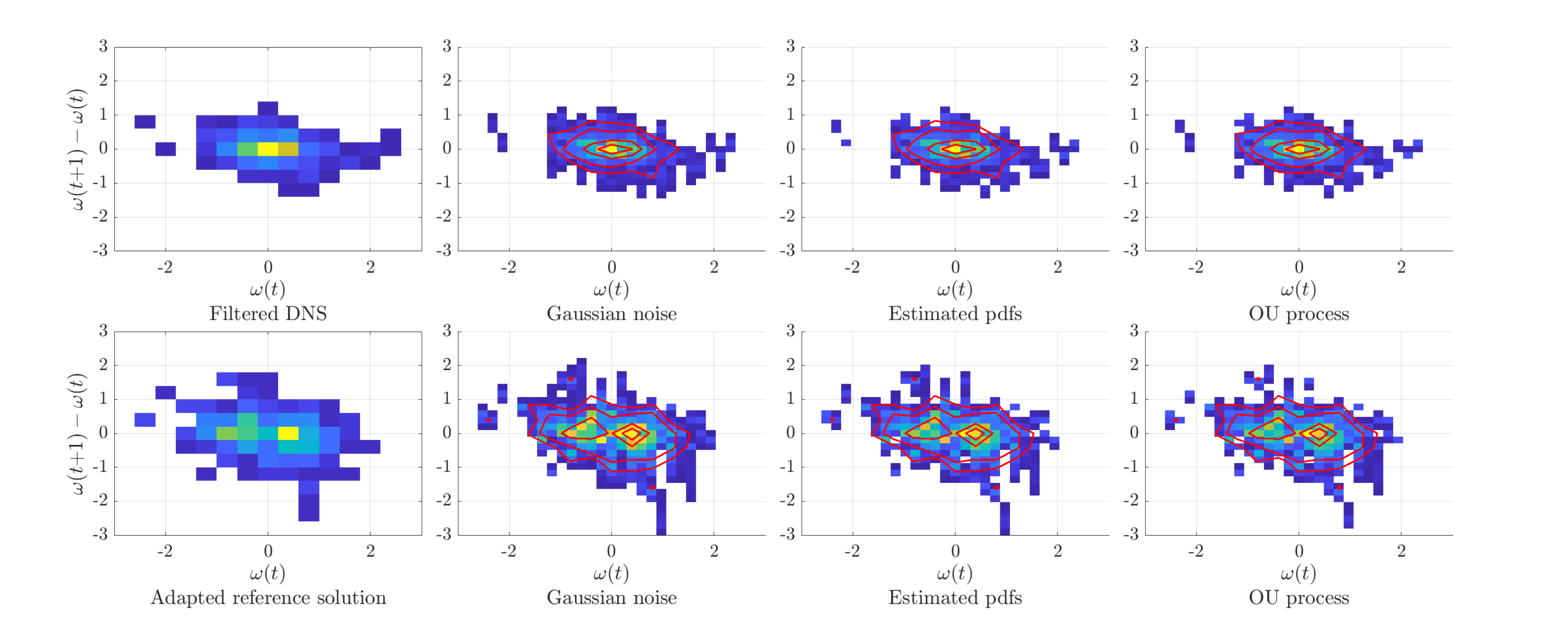}
    \caption{Conditional probability \eqref{eq:conditional_probability} for lead time $\tau=1$. The top row shows the distributions using the filtered DNS as a reference, the bottom row uses the adapted reference solution. The contour lines of the reference conditional distributions are overlaid on the distributions obtained from the stochastic ensembles for easier qualitative comparison.}
    \label{fig:condpdf_tau25}
\end{figure}

\begin{figure}[h!]
    \centering
    \begin{subfigure}[t]{0.45\textwidth}
    \includegraphics[width=\textwidth]{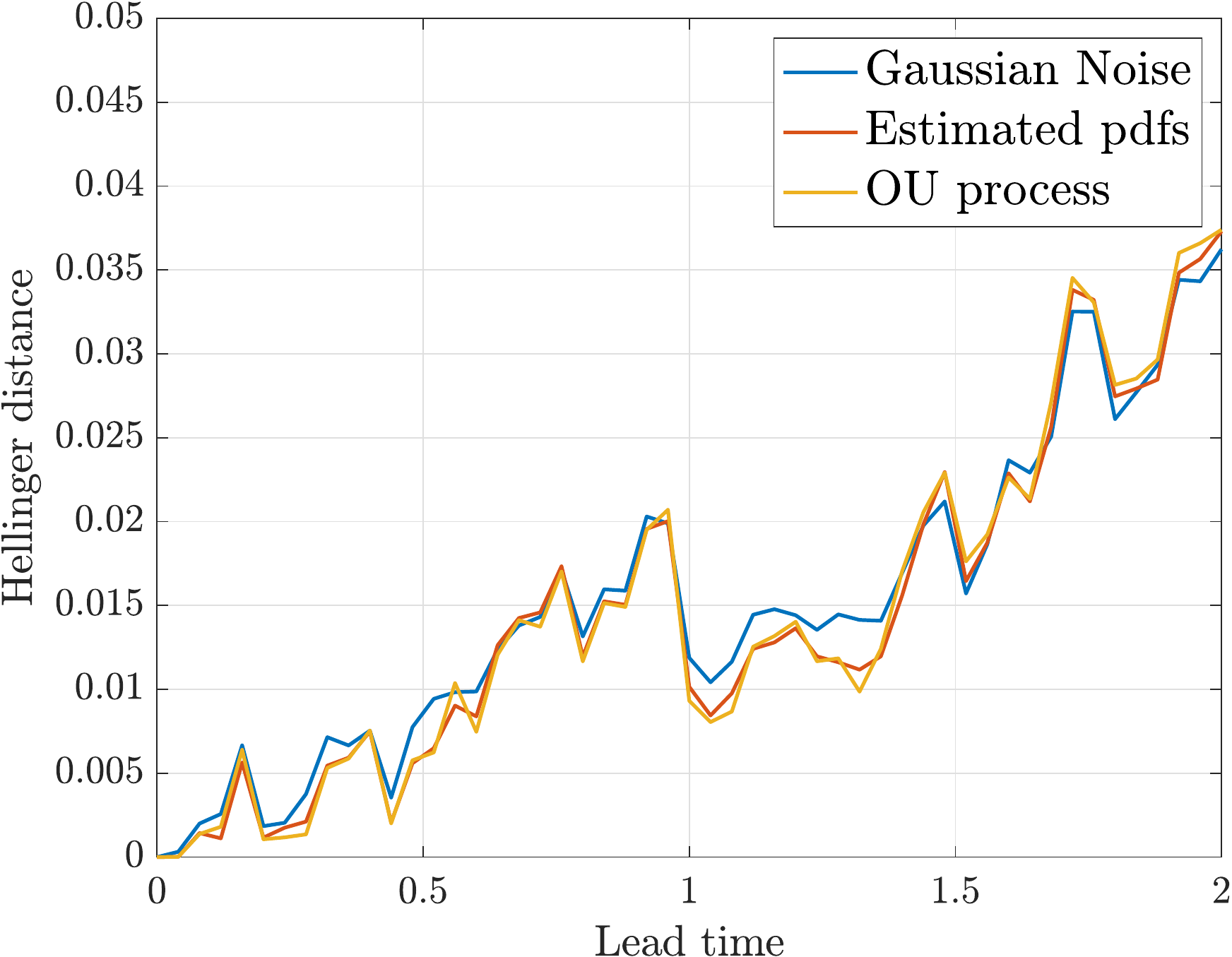}
    \end{subfigure}
    \begin{subfigure}[t]{0.45\textwidth}
    \centering
    \includegraphics[width=\textwidth]{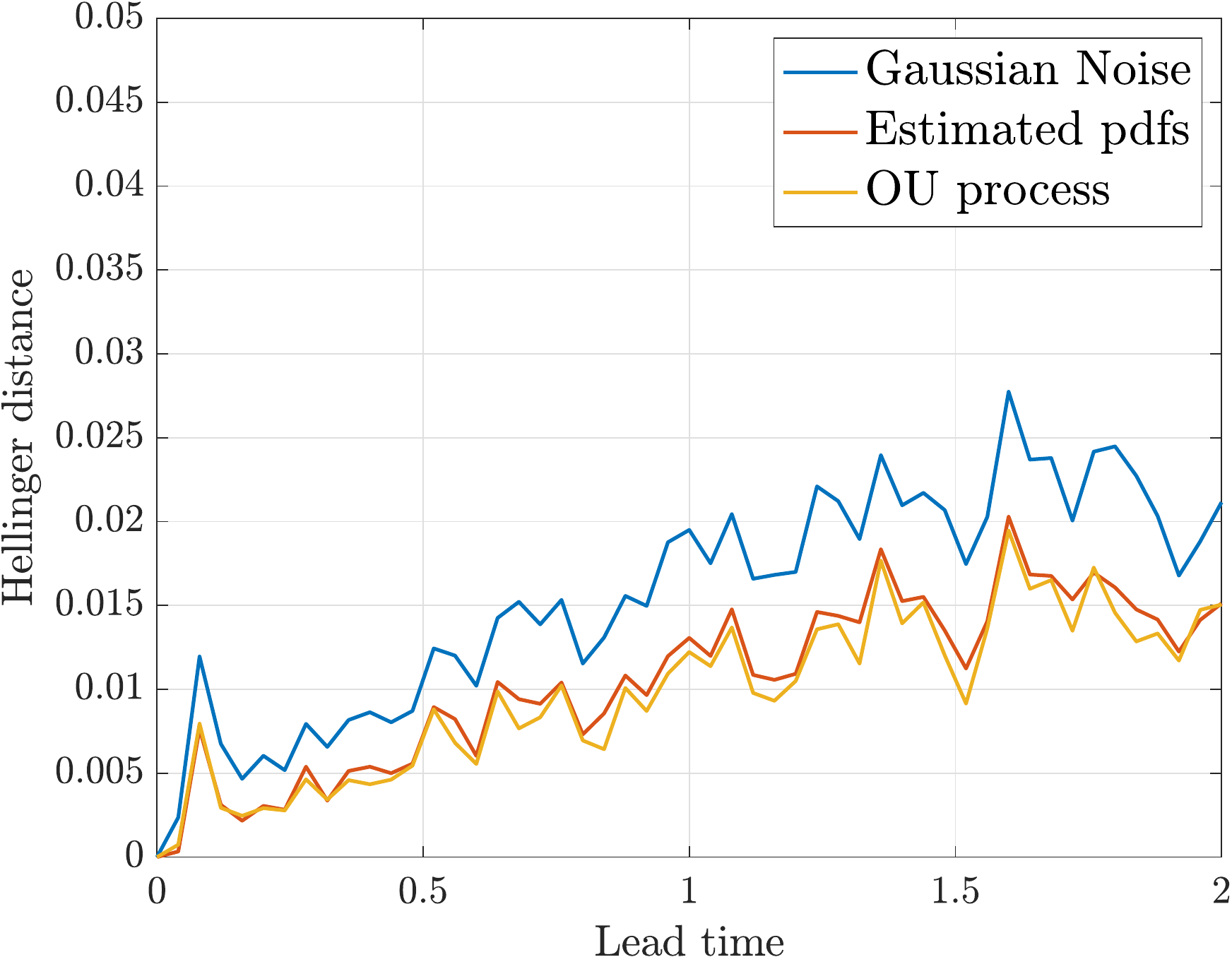}
    \end{subfigure}
    \caption{Hellinger distances as a function of time between the reference solution and the stochastic ensembles of distribution \eqref{eq:conditional_probability}. On the left, the filtered DNS is used as a reference solution, on the right, the adapted reference solution provides the reference.}
    \label{fig:Hellinger}
\end{figure}

The conditional distributions are shown in Fig. \ref{fig:condpdf_tau1}, at lead time $\tau=0.04$, and in Fig. \ref{fig:condpdf_tau25}, at lead time $\tau = 1$ to illustrate both short-time and long-time evolution. In both figures, the conditional distributions obtained from the reference solutions are shown in the left panel. For comparison, contour lines of these distributions have been overlaid in the conditional distributions obtained from the stochastic models. The filtered DNS provides the reference for the top row of distributions, the adapted reference solution is used in the bottom row. In particular, the distributions of the stochastic models have been computed from a set of initial conditions sampled along the filtered DNS and the adapted reference solution, respectively. In these figures, a large spread in the vertical direction indicates large uncertainty.  This becomes especially clear for the shortest lead times considered. On such short timescales, the stochastic forcing adds considerable variance to the numerical solution. Applying Gaussian noise yields the largest spread, whereas using the estimated pdfs and the OU produce a smaller spread, in accordance with previously presented results. At lead time $\tau=1$ (Fig. \ref{fig:condpdf_tau25}), the stochastic conditional distributions do not show significant differences.  To better judge the agreement between the stochastic conditional distributions and the reference distributions, we compute the Hellinger distance. This measure allows for a quantitative comparison between the different distributions. Given two discrete probability distributions $p=(p_1,\ldots,p_K)$ and $q=(q_1,\ldots,q_K)$, we compute the Hellinger distance \cite{hellinger1909neue} \begin{equation}
        H^2(p,q) = \frac{1}{2}\sum_{i=1}^K \left(\sqrt{p_i}-\sqrt{q_i}\right)^2. \label{eq:Hellinger}
    \end{equation}
The distance $H^2(p,q)$ of \eqref{eq:conditional_probability} is shown in Fig. \ref{fig:Hellinger} for the filtered DNS (left figure) and for the adapted solution (right figure). The initial conditions of the stochastic ensemble and the reference solutions are the same, therefore the Hellinger distance at $\tau=0$ is zero. As $\tau$ increases, $\omega(t)$ deviates from its reference value and accumulation of error leads to larger values of $H^2(p,q)$. Using the filtered DNS as reference solution yields a comparable Hellinger distance for each method. In contrast, the comparison of the stochastic ensembles to the adapted reference solution clearly favours the models obtained using the estimated pdfs and OU processes over those where Gaussian noise is employed. Despite the quantitative difference in the Hellinger distance, the qualitative behaviour is the same for each of the stochastic models.

An overall smaller rate of increase is observed when comparing to the adapted reference solution with respect to the filtered DNS. The latter findings underpin once more the benefits of using the adapted reference solution when assessing the quality of different stochastic models. 


\newpage
\section{\label{sec:conclusions}Conclusions}
In this paper, we have assessed three stochastic models for the simulation of the coarse-grained two-dimensional Euler equations. The closure is based on the so-called Stochastic Advection by Lie Transport (SALT) approach. The resulting SPDE contains a stochastic forcing term which requires to be modelled in order to close the equations. In particular, the forcing is decomposed into a deterministic basis (empirical orthogonal functions, or EOFs) multiplied by stochastic temporal traces. This decomposition is, by construction, fully determined from a fine-grid (DNS) dataset. However, to simulate outside the available dataset modeling the time traces is required. In the framework of SALT \cite{cotter2019numerically} the latter are regarded as Gaussian processes. Here we extend the stochastic forcing to more general processes, sampling from the data-estimated probability distribution functions (pdfs) and introducing correlation through Ornstein-Uhlenbeck (OU) processes. The latter two methods use additional data already available from the EOF time series. Between the methods no qualitative differences in the flow realizations were observed. However, the latter methods generally show favorable results compared to the former Gaussian method, in terms of ensemble mean and ensemble spread. 


In order to meaningfully compare the different stochastic models we defined a maximal prediction horizon and an adapted reference solution.
The prediction horizon sets the point in time beyond which a bundle of fine-grid solutions, starting from the same initial condition on the coarse grid, deviates on order 1 due to high sensitivity to the initial conditions. This defines the time frame on which to assess the statistical quality of the coarse-grid predictions. The adapted reference solution was defined as the coarse-grid solution using the exact measured time series of the EOFs for the forcing. The latter allowed to isolate the modelling error from other sources of error not taken into account in the considered model formulation, such as discretization error. The stochastic ensembles were compared to this reference solution using a global measure and pointwise values. For both the global and local measures, using either estimated pdfs or OU processes to define the forcing term yielded a smaller ensemble mean error and a smaller spread compared to using Gaussian noise.

Stochastic prediction ensembles on timescales relevant for data assimilation were further investigated by performing statistical tests, comparing ensembles of stochastic realizations to the adapted reference solution and the filtered DNS. A significantly smaller ensemble spread was found when using estimated pdfs or OU processes, compared to using Gaussian noise. Additionally, the observed mean ensemble error was lower for the former two methods. All three methods showed a rapid growth in ensemble error when compared to the filtered DNS, suggesting that the filtered DNS contains not only the modelling error but also the the discretization error and the closure error. These results were further substantiated by rank histograms, showing that the ensembles were biased with respect to the filtered DNS, but were underdispersive compared to the adapted reference solution. In particular, using the estimated pdfs to define the stochastic forcing rarely resulted in the adapted reference solution not being contained in the ensemble. Finally, conditional distributions of the vorticity were computed and compared using the Hellinger distance. Here, using estimated pdfs or OU processes resulted in a smaller distance to the reference solution than using Gaussian noise, indicating a better statistical characterization of the vorticity dynamics.

The methods presented in this paper may be used in other flows where EOF-based stochastic modeling is relevant. These approaches are particularly appealing since all information used in these methods is readily available from the EOF decomposition and no additional data is required to construct the models. The presented techniques are purely data-driven, they require no further assumption about the governing equations and can therefore be applied to other geophysical fluids. The short-time results indicate that a mean error reduction and smaller ensemble spread can be obtained using these methods, which can complement methods employed in data assimilation. Furthermore, the definition of the adapted reference solution motivates further research of the SALT method using different closure models and incorporating the discretization error.

\section*{Acknowledgements}
The authors would like to thank Wei Pan, at the Department of Mathematics, Imperial College London, for his help preparing the numerical experiments. We are grateful to thank Darryl Holm and James-Michael Leahy, at the Department of Mathematics, Imperial College London, and Arnout Franken, at the University of Twente, for the many inspiring discussions we had in the context of the SPRESTO project, funded by the Dutch Science Foundation (NWO) in their TOP1 program.

\bibliographystyle{apalike}
\bibliography{references}

\end{document}